\documentclass[twocolumn,showpacs,preprintnumbers,amsmath,amssymb,nofootinbib]{revtex4}

\usepackage{graphicx}
\usepackage{dcolumn}
\usepackage{bm}
\usepackage{amssymb}
\usepackage{amsmath}
\usepackage{amsthm}

\begin{document}


\title{Interference in spectrum of radiation from doubly scattered charged particle}

\author{M.V.~Bondarenco}
 \email{bon@kipt.kharkov.ua}
\author{N.F.~Shul'ga}
 \affiliation{NSC Kharkov Institute of Physics and Technology, 1 Academic Street,
61108 Kharkov, Ukraine}
 \affiliation{
V.N. Karazine Kharkov National University, 4 Svobody Square, 61077
Kharkov, Ukraine}

\date{\today}

\begin{abstract}

Existence of different types of interference in the spectrum of
radiation emitted by a doubly hard scattered electron is
demonstrated. The spectrum develops oscillations in two regions: the
hard, where the oscillations depend on the electron Lorentz factor,
and the soft, where the oscillations depend on the electron
scattering angles. This interference pattern owes to the presence of
jetlike radiation configurations, formed by a
piecewise-rectilinearly moving electron and the accompanying photon.
The corresponding nondipole decomposition relation is derived.
Notions describing proper field formation and interference, and
presumably being applicable more generally, are discussed in detail.


\end{abstract}





\maketitle

\section{Introduction}

Relatively recently, it was recognized that gamma radiation from
ultrarelativistic electrons in noncrystalline finite targets can
exhibit salient ``nondipole" interference effects when the electron
deflection angles exceed the natural scale for radiation emission
angles set by the electron inverse Lorentz factor
\cite{Blankenbecler,Zakharov,BK-structured,Bondarenco-Shulga}. The
existence of such effects was confirmed by CERN experiments, which
observed oscillations in the hard part of angle-integral photon
emission spectra from $\sim200$ GeV electrons passing through a pair
of amorphous foils separated by a submillimeter gap
\cite{NA63-plans}. Understanding of the behavior of radiation spectra in such cases, however, still seems to be incomplete. Reasoning of
\cite{Bondarenco-Shulga} had explained the shape and location of
features in the hard spectral domain, whereas the soft domain was
assumed to be featureless. But that holds only for spectra
averaged over a broad scattering angle distribution like that in
amorphous foils in experiments \cite{NA63-plans}.

Qualitative assessment of any interference effects in radiation
relies on the notion of photon formation length, which is
confronted with intrinsic geometrical scale(s) in the problem (in
the case mentioned above -- with the interfoil distance). The
conventionally defined photon formation length
\begin{equation}\label{coh-length-theta}
l_{\text{f}}=\frac{2}{\omega\left(\gamma^{-2}+\theta^2\right)},
\end{equation}
besides the photon frequency $\omega$, depends on its emission
angles $\theta$, and through them, indirectly, on electron
scattering angles. Thus, in practice, one has first to accurately determine
which angles are relevant, and from which particular direction they
are to be counted off. For instance, the Landau-Pomeranchuk-Migdal
suppression of the soft part of radiation spectrum from an electron
in a thick amorphous target \cite{LPM,Migdal} is known to be
described by the photon formation length depending on the mean
square deflection angle accumulated by the electron in the target,
with $\theta^2\sim\left\langle\chi^2\right\rangle\gg\gamma^{-2}$
\cite{Galitsky-Gurevich,BLP,Klein}. On the contrary, the
aforementioned case of nondipole bremsstrahlung on a pair of
amorphous foils \cite{NA63-plans} seems to be related only with the
\emph{free} photon formation length (defined in the absence of
electron scattering), when $\theta\lesssim\gamma^{-1}$
\cite{Bondarenco-Shulga}. It may be puzzling how to reconcile this
with the fact that the corresponding oscillations fully develop only
when typical scattering angles overtake the inverse Lorentz factor.
Basically, that can be due to incidental insensitivity of period and phase of
the oscillations in the hard spectral domain to the mean square
angles of scattering in both targets, but a principal question remains: Is
there a signature of the electron scattering angle dependent photon
formation length anywhere among the radiation observables?

To answer this question, and pave the way for further developments,
it is expedient to revisit the cornerstone problem -- radiation from
an electron undergoing successive double scattering through certain
angles, not subject to any averaging. Such a problem was discussed
in a number of instances before: The space-time evolution of the
retarded electromagnetic field was analyzed by Purcell
\cite{Purcell}, whereas general properties of the quantum amplitude,
by Feynberg \cite{Feynberg}.

The object of our study, however, will be the radiation spectrum
integrated over emission angles, to which experimental observation in the ultrarelativistic case is usually restricted, as long as photons are typically emitted in a narrow
cone around the forward direction. For this observable, the
interference pattern appears to be richer than one might naively
expect, and effectively involves manifestations of several photon
formation lengths, showing up in different spectral regions. We will
deduce the corresponding decomposition relation, and investigate
the physical meaning of its entries.

Examination of the emerging structure then leads us to more
profound conclusions. All the discovered spectral features prove to be consequences of the presence of electron-photon jetlike configurations,
which can participate in interference phenomena in spite of their
narrow collimation properties. That gives rise
to notions such as intermediate electron line and ``radio'"
contributions (among which the first is independent of electron
deflection angles, while the second is independent of the electron
Lorentz factor), and ``proper field form factors" multiplying the
radio contribution. The latter notions may be applicable in a broad class of problems. To reach their versatile understanding, it appears
beneficial to discuss the problem from
several points of view. 

Specifically, after setting forth the initial assumptions in
Sec.~\ref{sec:definitions}, we turn in Sec.~\ref{subsec:2.1} to
evaluating the spectrum in terms of photon emission angles. That
reveals the existence of jetlike and interjet radiation, but hides their spatial ordering. Additional insight is gained in
Sec.~\ref{subsec:imp-par} by considering the process in transverse
coordinates (impact parameters), which best elucidates the origin of
electron proper field form factors. In Sec.~\ref{sec:double-time},
that is complemented by a study of longitudinal evolution of photon
formation and interference, aiming to demonstrate that the long and
short scales anticipated to be photon formation lengths do correspond
to the process development in time. Section~\ref{sec:feasibility}
includes a brief analysis of experimental realizability of the
considered process. Section~\ref{sec:summary} provides the summary.
In the Appendix, we derive the covariant form of the double time
integral representation for the radiation spectrum, and highlight its
gauge properties, which prove relevant in the present context.

\section{Preliminaries}\label{sec:definitions}

For the interference effects discussed in this paper to be
pronounced, energies of the emitted photons are to be low compared
with the electron energy. That creates premises for applicability of
classical electrodynamics: The possible quantal nature of the
electron motion in the domains of scattering is inessential provided
the photon formation length greatly exceeds the extent of each of
those scattering areas. Then, the factorization theorem asserts that
the differential probability of the entire bremsstrahlung process
splits into a product of two differential cross sections of elastic
electron scattering and the differential probability of emission of
an electromagnetic wave from an angle-shaped charged particle
trajectory \cite{BLP,factorization-theorem}. Our study in this paper
will focus on the photon emission probability alone, which is
tractable purely classically. Moreover, at high energies, the motion
of the electron in macroscopic-field deflectors may be
semiclassical, as well (see Sec.~\ref{sec:feasibility}).

We thus consider radiation from a classical charge $e$ (physically
representing an electron or positron) moving along a
double-angle-shaped trajectory $\bm{r}(t)$, with velocity
$\bm{v}(t)=d\bm{r}/dt$ depending on time $t$. Specifically, as was
mentioned in the Introduction, we shall evaluate the angle-integral
radiation spectrum, which in the ultrarelativistic case
$\gamma=(1-v^2)^{-1/2}\ggg1$ is the prime experimental observable.
It involves several operations \cite{Jackson}: a time integral with
conjugate plane-wave factor $e^{i\omega t-i\bm{k}\cdot\bm{r}}$
depending on the photon frequency $\omega$ and emission direction
$\bm{n}=\bm{k}/\omega$, subsequent amplitude squaring, and
integration over directions of $\bm{n}$:
\begin{equation}\label{dIdomega-through-amp}
\frac{dI}{d\omega}=\omega^2\int d^2n
\left|\frac{e}{2\pi}\int_{-\infty}^{\infty} dt
[\bm{n}\times\bm{v}(t)]e^{i\omega t-i\bm{k}\cdot\bm{r}(t)}\right|^2.
\end{equation}
To reach proper understanding of its behavior, it is desirable to
reduce (\ref{dIdomega-through-amp}) at least to a single integral,
enabling clear-cut isolation of dominant contributions, and thereby,
a rigorous measure of the radiation coherence. The physical meaning
of the latter contributions will depend on the nature of the last
integration variable. We will describe three most informative
approaches, and demonstrate that although, inevitably, they all lead
to the same structure of the final result, their interpretations
elucidate different aspects of the radiation process, thus being
mutually complementary. Their amalgamation then leads to a cogent
picture for interference effects in bremsstrahlung at electron
rescattering, which may also prove relevant for in other highly
nondipole radiation problems.

\section{Analysis in terms of angular distributions}\label{subsec:2.1}

\begin{figure}
\includegraphics{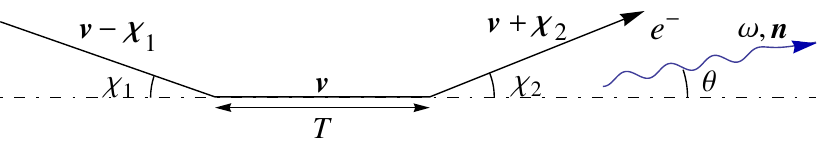}
 \caption{\label{fig:traj-scheme} Geometry of the considered electron scattering process and the accompanying radiation.
 For symmetry's sake, all the angles are counted off from the intermediate electron velocity.}
\end{figure}

To fully describe the electron trajectory, we denote its successive
elastic deflection angles as $\bm{\chi}_1$ and $\bm{\chi}_2$
[$\gamma^{-1}\ll \chi_{1,2}\ll1$], and the separation time (or
length)\footnote{We adopt units in which the speed of light equals
unity.} between the scatterings as $T$ (see
Fig.~\ref{fig:traj-scheme}).
Noncoplanarity of the scattering angles is characterized by the
azimuthal angle between them, $\varphi_{12}=\arccos
\left(\bm{\chi}_1\cdot\bm{\chi}_2/|\bm{\chi}_1||\bm{\chi}_2|\right)$.
In this abrupt scattering case, it is straightforward to integrate
in (\ref{dIdomega-through-amp}) first over time. That gives
\begin{eqnarray}\label{dIdomega=intd2theta}
\frac{dI}{d\omega}
=\frac{e^2}{\pi^2}\int d^2\theta\Bigg\{\left[\frac{\bm{\theta}}{\gamma^{-2}+\theta^2}-\frac{\bm{\theta}+\bm{\chi}_1}{\gamma^{-2}+(\bm{\theta}+\bm{\chi}_1)^2}\right]^2\quad\nonumber\\
+\left[\frac{\bm{\theta}-\bm{\chi}_2}{\gamma^{-2}+(\bm{\theta}-\bm{\chi}_2)^2}-\frac{\bm{\theta}}{\gamma^{-2}+\theta^2}\right]^2\quad
\nonumber\\
+2\left[\frac{\bm{\theta}}{\gamma^{-2}+\theta^2}-\frac{\bm{\theta}+\bm{\chi}_1}{\gamma^{-2}+(\bm{\theta}+\bm{\chi}_1)^2}\right]\qquad\qquad\qquad\quad \nonumber\\
\cdot
\left[\frac{\bm{\theta}-\bm{\chi}_2}{\gamma^{-2}+(\bm{\theta}-\bm{\chi}_2)^2}-\frac{\bm{\theta}}{\gamma^{-2}+\theta^2}\right]\cos\frac{\omega
T}{2\gamma^2}\left(1+\gamma^2\theta^2\right)\Bigg\}.
\end{eqnarray}

\begin{figure}
\includegraphics{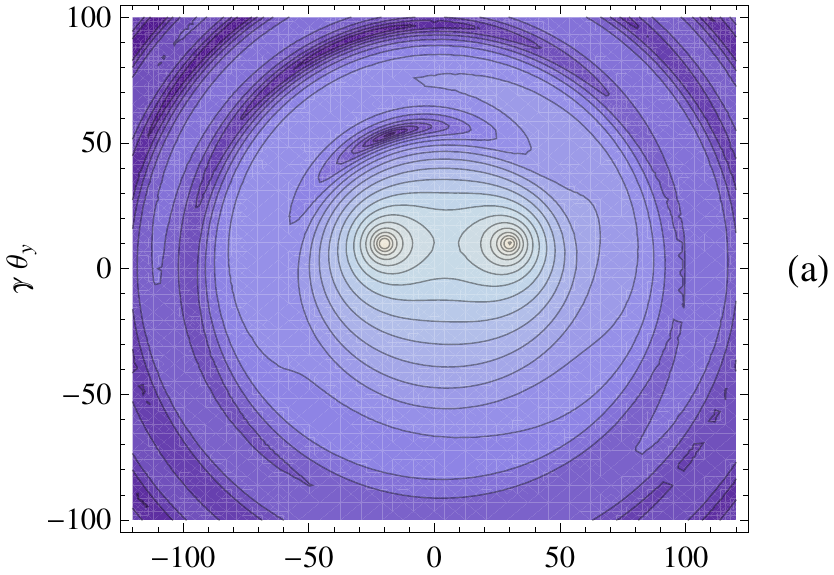}
\includegraphics{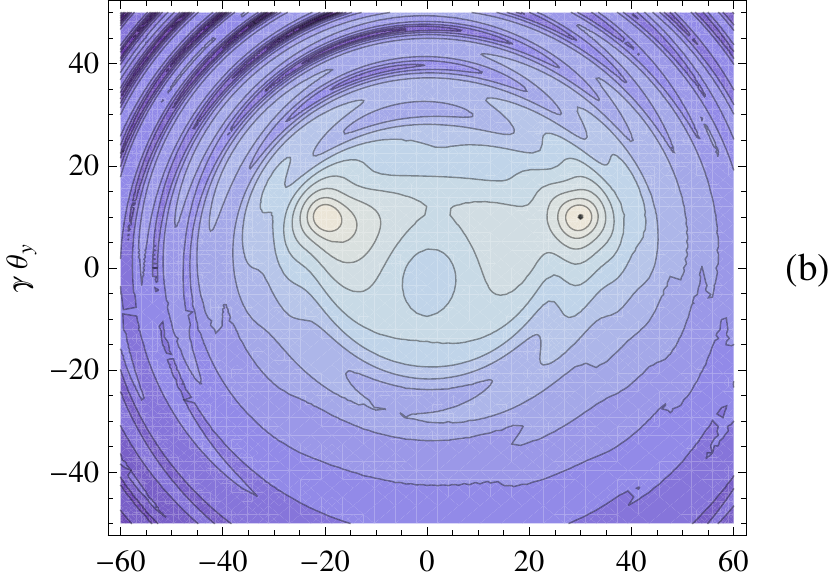}
\includegraphics{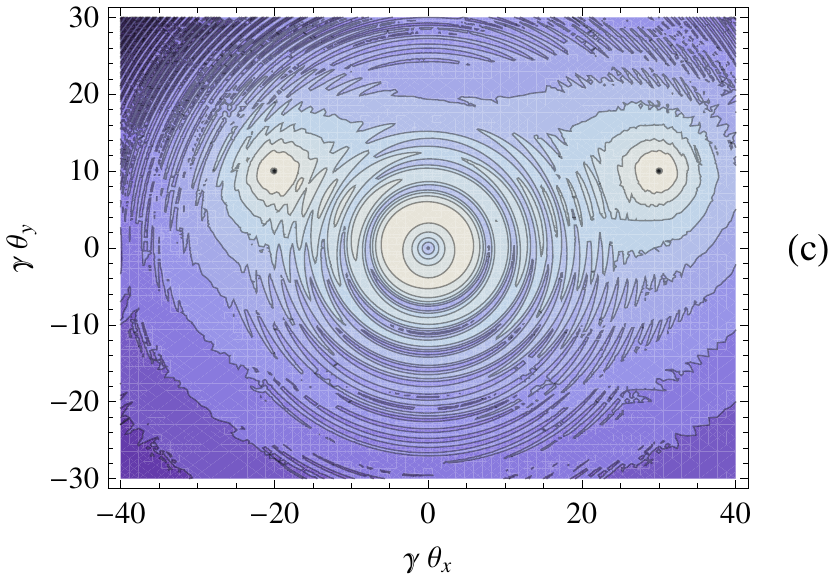}
 \caption{\label{fig:Ang} Angular distributions of radiation from a doubly scattered electron.
 Directions of initial and final electron motion coincide with centers of the leftmost and the rightmost jetlike features.
 (a) $\frac{\omega T}{2\gamma^2}=0.001$, (b) $\frac{\omega T}{2\gamma^2}=0.01$, 
(c) $\frac{\omega T}{2\gamma^2}=0.1$. For higher $\omega$, see
Fig.~\ref{fig:Ang1d-semibare} below. For discussion see text.}
\end{figure}

\begin{figure}
\includegraphics{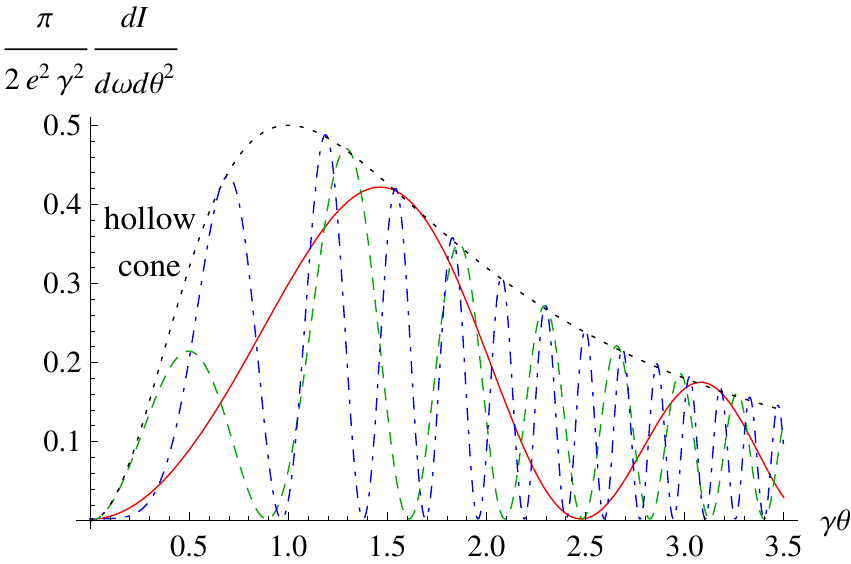}
 \caption{\label{fig:Ang1d-semibare} Angular distribution of radiation intensity around the direction of the intermediate electron velocity, for $\chi_1=\chi_2=30\gamma^{-1}$ (there is no sensitivity to those angles, provided they are large enough), and different $\omega$ in the hard region. In the displayed domain, the distribution is virtually axially symmetric and $\chi$-independent.
 Red solid curve, $\frac{\omega T}{2\gamma^2}=0.9$ (the main maximum of $dI/d\omega$ in the hard region). Green dashed, $\frac{\omega T}{2\gamma^2}=3.5$ (the following minimum of $dI/d\omega$).
 Blue dot-dashed, $\frac{\omega T}{2\gamma^2}=6.5$ (the secondary maximum of $dI/d\omega$). For the corresponding features in $dI/d\omega$, cf. Fig.~\ref{fig:Oscill}(a) below. Black dotted curve, the envelope $\frac{2\gamma^2\theta^2}{(1+\gamma^2\theta^2)^2}$.}
\end{figure}

The behavior of the integrand of Eq.~(\ref{dIdomega=intd2theta}) in
the $\bm{\theta}$ plane (the angular distribution of radiation) is
depicted in Figs.~\ref{fig:Ang} and \ref{fig:Ang1d-semibare} for a
case of exemplary scattering angles
$|\bm{\chi}_1|\sim|\bm{\chi}_2|\sim30\gamma^{-1}$, and progressively
increasing values of $\frac{\omega T}{2\gamma^2}$. One can visualize
there three cones of radiation (associated with one internal and two
external electron lines), with concentric interference rings about
the internal line. The outreach of the latter rings depends on
$\omega T$. At $\omega T\chi^2/2\to0$, the rings expand to infinity,
and the angular distribution of radiation at finite $\theta$ [the
inner part of Fig.~\ref{fig:Ang}(a)] tends to that at single
electron scattering through angle $\bm{\chi}_1+\bm{\chi}_2$ (cf.,
e.g., \cite{Fomin}).\footnote{It should be mentioned that Fig.~1 in
\cite{Fomin}, corresponding to the total deflection angle
$10\gamma^{-1}$, did not belong to the ultranondipole regime
yet, whereas our figure with $|\bm{\chi}_1+\bm{\chi}_2|\sim
50\gamma^{-1}$ does. That is responsible for residual differences
between the plots.} At some finite $\omega$, the radius of the rings
starts to come close to one of the deflection angles. Successively, when
this radius by far exceeds $\chi_1$, $\chi_2$, there exists only
interference between the external lines [Fig.~\ref{fig:Ang}(a)],
when it becomes on a par with the size of (one of the) $\chi$'s,
there emerges interference between the internal and an external line
[Fig.~\ref{fig:Ang}(b)], and when it falls below $\chi_1$, $\chi_2$,
the latter interference is lost, as well [Fig.~\ref{fig:Ang}(c)],
and only interference within the internal line survives
(Fig.~\ref{fig:Ang1d-semibare}).\footnote{Angular distributions
similar to those in Fig.~\ref{fig:Ang1d-semibare} were discovered
previously in other but related physical problems: radiation from an
electron in a straight section between magnets in a storage ring, or
in more elaborate synchrotron radiation sources (see
\cite{synchr-rad-straight-section,Geloni} and references therein).}

For arbitrary $\bm{\chi}_1$ and $\bm{\chi}_2$, the
$\omega$-independent (noninterference) part of
(\ref{dIdomega=intd2theta}) consists of two separate
(``Bethe-Heitler") contributions from the scattering vertices, each
of which is given by the well-known bremsstrahlung formula
\begin{subequations}
\begin{eqnarray}\label{I0}
\frac{dI_{\text{BH}}}{d\omega}(\gamma\chi)\!&=&\!\frac{e^2}{\pi^2}\int d^2\theta\left[\frac{\bm{\theta}-\bm{\chi}}{\gamma^{-2}+(\bm{\theta}-\bm{\chi})^2}-\frac{\bm{\theta}}{\gamma^{-2}+\theta^2}\right]^2\nonumber\\
&\,&\label{I0-int}\\
&\underset{\chi\gg\gamma^{-1}}\simeq&
\frac{2e^2}{\pi}\left(\ln\gamma^2\chi^2-1\right).
\end{eqnarray}
\end{subequations}
Integral (\ref{I0-int}) converges due to mutual cancellation between
the terms in the brackets.

In contrast, for the nontrivial, interference part of
(\ref{dIdomega=intd2theta}), the terms in its integrand may be
treated independently, because there the convergence is provided by
the cosine factor. The key observation is that different terms in
the integrand give dependence of $dI/d\omega$ on $\omega$ at
different scales. Specifically, separating $\chi$-dependent and
-independent parts gives
\begin{eqnarray}\label{dIdomega-separ-theta-terms}
\frac{dI}{d\omega}
&=&\frac{dI_{\text{BH}}}{d\omega}(\gamma\chi_1)+\frac{dI_{\text{BH}}}{d\omega}(\gamma\chi_2)\nonumber\\
&\,&+\frac{2e^2}{\pi}\left[g\left(\frac{\omega
T}{2\gamma^2}\right)+r\left(\bm{\chi}_1,\bm{\chi}_2,\gamma^{-1},\omega
T\right)\right],
\end{eqnarray}
where\footnote{In \cite{Bondarenco-Shulga}, function
$g\left(\frac{\omega T}{2\gamma^2}\right)$ was denoted as
$g_{\text{ll}}\left(0,\frac{\omega T}{2\gamma^2}\right)$, to
distinguish it from dipole or mixed-strength radiation cases. Here,
confining ourselves to the strongly nondipole case alone, we omit
noninformative labels.}
\begin{subequations}\label{g-def}
\begin{eqnarray}
g\left(\frac{\omega T}{2\gamma^2}\right)&=&-\frac1{\pi}\int
d^2\theta\frac{\theta^2}{(\gamma^{-2}+\theta^2)^2}\cos\frac{\omega
T}{2\gamma^2}\left(1+\gamma^2\theta^2\right)\nonumber\\
&\,&\label{gll-integral}\\
&=&\text{Ci}\left(\frac{\omega T}{2\gamma^2}\right)+\frac{\omega
T}{2\gamma^2}\text{si}\left(\frac{\omega
T}{2\gamma^2}\right)+\cos\frac{\omega T}{2\gamma^2},\label{Ci+si}
\end{eqnarray}
\end{subequations}
with $\text{Ci}(z)=-\int_z^{\infty}\frac{dx}{x}\cos x$,
$\text{si}(z)=-\int_z^{\infty}\frac{dx}{x}\sin x$ the integral cosine and sine functions \cite{Abr-Stegun},
and
\begin{eqnarray}\label{h-def}
r\left(\bm{\chi}_1,\bm{\chi}_2,\gamma^{-1},\omega
T\right)=\frac{1}{\pi}\int d^2\theta \cos\frac{\omega
T}{2\gamma^2}\left(1+\gamma^2\theta^2\right)\quad\nonumber\\
\times\Bigg\{\frac{\bm{\theta}}{\gamma^{-2}+\theta^2}\cdot \left[\frac{\bm{\theta}+\bm{\chi}_1}{\gamma^{-2}+(\bm{\theta}+\bm{\chi}_1)^2}+\frac{\bm{\theta}-\bm{\chi}_2}{\gamma^{-2}+(\bm{\theta}-\bm{\chi}_2)^2}\right]\nonumber\\
-\frac{\bm{\theta}+\bm{\chi}_1}{\gamma^{-2}+(\bm{\theta}+\bm{\chi}_1)^2}
\cdot
\frac{\bm{\theta}-\bm{\chi}_2}{\gamma^{-2}+(\bm{\theta}-\bm{\chi}_2)^2}
\Bigg\}.
\end{eqnarray}
Note that $\int_0^{\infty}d\omega g\left(\frac{\omega
T}{2\gamma^2}\right)=\int_0^{\infty}d\omega
r\left(\bm{\chi}_1,\bm{\chi}_2,\gamma^{-1},\omega T\right)=0, $ as a
consequence of locality of electromagnetic energy emission in
classical electrodynamics (see, e.g., \cite{Bondarenco-Shulga}).

At $\omega T/2\gamma^2\gtrsim 1$, when typical contributing angles
are restricted by $\theta\lesssim\gamma^{-1}\ll\chi$, part $r$ is suppressed compared to $g$ by inverse powers of $\gamma\chi$, and can be neglected. Therewith,
\begin{equation}\label{dIdomega-highomega}
\frac{dI}{d\omega} \underset{\frac{\omega T}{2\gamma^2}\gtrsim
1}\simeq
\frac{dI_{\text{BH}}}{d\omega}(\gamma\chi_1)+\frac{dI_{\text{BH}}}{d\omega}(\gamma\chi_2)+\frac{2e^2}{\pi}g\left(\frac{\omega
T}{2\gamma^2}\right).
\end{equation}
In the formal limit $\omega T/2\gamma^2\to\infty$, the spectrum
exhibits decreasing harmonic oscillations \cite{Bondarenco-Shulga}
\footnote{Strictly speaking, (\ref{2I0+cos}) becomes numerically
accurate only in a rather far asymptotic region (see
\cite{Bondarenco-Shulga}). To reach higher precision, it may be
worth retaining the next-to-leading order term in the phase, but we
shall not indulge into such complications in the present paper.}
\begin{equation}\label{2I0+cos}
\frac{dI}{d\omega} \underset{\frac{\omega T}{2\gamma^2}\gg 1}\simeq
\frac{dI_{\text{BH}}}{d\omega}(\gamma\chi_1)+\frac{dI_{\text{BH}}}{d\omega}(\gamma\chi_2)+\frac{2e^2}{\pi}\left(\frac{2\gamma^2}{\omega
T}\right)^2\cos\frac{T}{l_0(\omega)},
\end{equation}
where
\begin{equation}\label{lf-free}
l_0(\omega)=\frac{2\gamma^2}{\omega}
\end{equation}
stands for the ``free" photon formation length. It is relevant here
insofar as between the hard scatterings the electron moves strictly
rectilinearly (should there be some medium or external field along
its path, the situation might drastically change). Note, too, that the
decrease here follows the law $\sim\omega^{-2}$ instead of
$\sim\omega^{-1}$, owing to the integrand in (\ref{gll-integral})
vanishing at $\theta\to0$ (a ``hollow cone" distribution of
radiation emitted from an isolated straight electron line), due to
the vector and gauge character of electromagnetic radiation.

When $\omega\to0$, function (\ref{Ci+si}) logarithmically diverges,
so, ultimately, approximation (\ref{dIdomega-highomega}) must break
down. That reflects physical limitedness of separate treatment of
individual terms in
Eq.~(\ref{dIdomega-separ-theta-terms}).\footnote{It must be
remembered that the interpretation of individual terms in
(\ref{dIdomega-through-amp}) as stemming from isolated parts of the
electron's trajectory is not gauge invariant, because at the ends of
a finite trajectory segment there is no conservation of charge. In
this and the following sections, we work, specifically, in the
radiative gauge. Nevertheless, since the entire spectral-angular
distribution (\ref{dIdomega=intd2theta}) is gauge invariant, its
terms with different dependencies on $\bm{\chi}_1$, $\bm{\chi}_2$
may be singled out at least formally, and treated separately in this
sense.} At sufficiently low $\omega$, all the terms in
(\ref{dIdomega-separ-theta-terms}) become comparable and
simultaneously important. Nonetheless, there still remains room for
simplifications: At $\chi\gg\gamma^{-1}$, it is justified to
entirely neglect terms containing $\gamma^{-2}$ (provided
$|\bm{\chi}_1+\bm{\chi}_2|\gg\gamma^{-1}$, to avoid a case of
overlap of singularities, which will be touched upon later).
Physically, that means that in the softest spectral (here called
radio) region, the ensemble of segments of the electron's trajectory
acts like a single antenna (cf., e.g., \cite{Dokshitzer}) -- viz.,
like a long ``wire", which is significantly deformed within the
photon formation length, so that the electric current along it,
representing the passing electron, may be regarded as flowing
exactly at the speed of light. Evaluation of the corresponding integral
gives\footnote{Integration over the azimuth of $\bm{\theta}$ is
alleviated by introducing a complex variable
$\zeta=\theta_x+i\theta_y$ for Cartesian components $\theta_x$,
$\theta_y$ of vector $\bm{\theta}$, and evaluating the encountered
integrals by residues:
\begin{eqnarray}\label{conf-int-1}
\int d\phi_{\bm{\theta}}\frac{\bm{\theta}\cdot(\bm{\theta}-\bm{\chi})}{(\bm{\theta}-\bm{\chi})^2}
=\mathfrak{Re}\int d\phi_{\bm{\theta}}\frac{(\theta_x+i\theta_y)(\theta_x-i\theta_y-\chi)}{(\theta_x+i\theta_y-\chi)(\theta_x-i\theta_y-\chi)}\nonumber\\
=\mathfrak{Re}\frac{1}{i}\oint_{|\zeta|=|\bm{\theta}|}\frac{d\zeta}{\zeta-\chi}=2\pi\vartheta\left(|\bm{\theta}|-|\bm{\chi}|\right),
\end{eqnarray}
with $\vartheta(\ldots)$ the Heaviside unit step function, and
\begin{eqnarray}\label{conf-int-2}
\int
d\phi_{\bm{\theta}}\frac{(\bm{\theta}+\bm{\chi}_1)\cdot(\bm{\theta}-\bm{\chi}_2)}{(\bm{\theta}+\bm{\chi}_1)^2(\bm{\theta}-\bm{\chi}_2)^2}
=\mathfrak{Re}\oint_{|\zeta|=|\bm{\theta}|} d\phi_{\zeta}\frac{1}{(\bar{\zeta}-\bar{\chi}_2)(\zeta+\chi_1)}\nonumber\\
=\mathfrak{Re}\frac1{\bar{\chi}_2\chi_1+|\zeta|^2}\oint_{|\zeta|=|\bm{\theta}|} d\phi_{\zeta}
\left[\frac{\bar{\zeta}}{\bar{\zeta}-\bar{\chi}_2}-\frac{\chi_1}{\zeta+\chi_1}\right]\nonumber\\
=2\pi\left[\vartheta\left(|\bm{\theta}|-|\bm{\chi}_2|\right)-\vartheta\left(|\bm{\chi}_1|-|\bm{\theta}|\right)\right]\mathfrak{Re}\frac1{\bar{\chi}_2\chi_1+\theta^2},
\end{eqnarray}
with $\chi_1=\chi_{1x}+i\chi_{1y}$, $\bar
\chi_2=\chi_{2x}-i\chi_{2y}$.}
\begin{subequations}\label{h-eval}
\begin{eqnarray}\label{h-single-int}
&\,&r\left(\bm{\chi}_1,\bm{\chi}_2,0,\omega
T\right)\nonumber\\
&\,&=\frac1{\pi}\int d^2\theta
\Bigg\{\frac{\bm{\theta}}{\theta^2}\cdot\left[\frac{\bm{\theta}+\bm{\chi}_1}{(\bm{\theta}+\bm{\chi}_1)^2}+\frac{\bm{\theta}-\bm{\chi}_2}{(\bm{\theta}-\bm{\chi}_2)^2}\right]\nonumber\\
&\,&- \frac{\bm{\theta}+\bm{\chi}_1}{(\bm{\theta}+\bm{\chi}_1)^2}\cdot\frac{\bm{\theta}-\bm{\chi}_2}{(\bm{\theta}-\bm{\chi}_2)^2}\Bigg\}\cos\frac{\omega T\theta^2}2\nonumber\\
&\,&=\int_{\chi_1^2}^{\infty}\frac{d\theta^2}{\theta^2}\cos\frac{\omega T\theta^2}{2}+\int_{\chi_2^2}^{\infty}\frac{d\theta^2}{\theta^2}\cos\frac{\omega T\theta^2}{2}
\nonumber\\
&\,&-
\int_{\chi_1^2}^{\infty}d\theta^2\cos\frac{\omega T\theta^2}{2}\mathfrak{Re}\frac1{\chi_1\bar{\chi}_2+\theta^2}\nonumber\\
&\,&-
\int_{\chi_2^2}^{\infty}d\theta^2\cos\frac{\omega T\theta^2}{2}\mathfrak{Re}\frac1{\chi_1\bar{\chi}_2+\theta^2}
\nonumber\\
&\,&+\int_{0}^{\infty}d\theta^2\cos\frac{\omega
T\theta^2}{2}\mathfrak{Re}\frac1{\chi_1\bar{\chi}_2+\theta^2},
\end{eqnarray}
where
$\chi_1\bar\chi_2=(\chi_{1x}+i\chi_{1y})(\chi_{2x}-i\chi_{2y})=|\bm{\chi}_1||\bm{\chi}_2|e^{i\varphi_{12}}$.
Representation (\ref{h-single-int}) in terms of single integrals is
already suitable for assessment of coherence effects, but those
integrals can be readily taken, as well:
\begin{eqnarray}\label{FA-def}
&\,&r\left(\bm{\chi}_1,\bm{\chi}_2,0,\omega
T\right)\nonumber\\
&\,&=-\text{Ci}\left(\frac{\omega T\chi_1^2}{2}\right)-\text{Ci}\left(\frac{\omega T\chi_2^2}{2}\right)\nonumber\\
&\,&+\mathfrak{Re}\Bigg\{\cos\frac{\omega T\chi_1\bar\chi_2}{2}\Bigg(\text{Ci}\left[\frac{\omega T}2(\chi_1^2+\chi_1\bar\chi_2)\right]\nonumber\\
&\,&+\text{Ci}\left[\frac{\omega T}2(\chi_2^2+\chi_1\bar\chi_2)\right]-\text{Ci}\left[\frac{\omega T}2\chi_1\bar\chi_2\right]\Bigg)\nonumber\\
&\,&+\sin\frac{\omega T\chi_1\bar\chi_2}{2}\Bigg(\text{si}\left[\frac{\omega T}2(\chi_1^2+\chi_1\bar\chi_2)\right]\nonumber\\
&\,&+\text{si}\left[\frac{\omega
T}2(\chi_2^2+\chi_1\bar\chi_2)\right]-\text{si}\left[\frac{\omega
T}2\chi_1\bar\chi_2\right]\Bigg)\Bigg\}.
\end{eqnarray}
\end{subequations}
This function must be added to (\ref{Ci+si}), with a proviso that
owing to the admitted neglect of $\gamma^{-1}$, the validity of
approximation (\ref{FA-def}) is restricted to the domain $\omega
T\lesssim \chi^{-2}$.

At $\gamma^{-2}\ll\omega T\ll\gamma\chi^{-1}$, approximation
(\ref{h-eval}) devolves to decreasing harmonic oscillations:
\begin{eqnarray}\label{asympt-sin-complex-form}
r\left(\bm{\chi}_1,\bm{\chi}_2,0,\omega T\right)\simeq\frac2{\omega T}\left(\mathfrak{Re}\frac{1}{\chi_1^2+\chi_1\bar\chi_2}-\frac1{\chi_1^2}\right)\sin\frac{\omega T\chi_1^2}2\,\nonumber\\
+\frac2{\omega
T}\left(\mathfrak{Re}\frac{1}{\chi_2^2+\chi_1\bar\chi_2}-\frac1{\chi_2^2}\right)\sin\frac{\omega
T\chi_2^2}2.\nonumber\\
&\,&
\end{eqnarray}
But their decrease rate appears to be too slow, so in the far
asymptotics, this approximation needs to be corrected. To this end,
instead of considering the full integrand in (\ref{h-def}), it
suffices to single out only its most singular parts -- vicinities of
points $\bm{\theta}=-\bm{\chi}_1$ and $\bm{\theta}=\bm{\chi}_2$. The calculation then gives
\begin{eqnarray}\label{interm-omega-K1}
r\left(\bm{\chi}_1,\bm{\chi}_2,\gamma^{-1},\omega T\right)\simeq \frac1{\pi}\left(\frac{\bm{\chi}_1+\bm{\chi}_2}{(\bm{\chi}_1+\bm{\chi}_2)^2}- \frac{\bm{\chi}_1}{\chi_1^2}\right)\qquad\qquad\nonumber\\
\cdot\int d^2\theta\frac{\bm{\theta}+\bm{\chi}_1}{\gamma^{-2}+(\bm{\theta}+\bm{\chi}_1)^2}\cos\frac{\omega T}{2}\left[\chi_1^2-2\bm{\chi}_1\cdot(\bm{\theta}+\bm{\chi}_1)\right]\nonumber\\
+\frac1{\pi}\left(\frac{\bm{\chi}_2}{\chi_2^2}-\frac{\bm{\chi}_1+\bm{\chi}_2}{(\bm{\chi}_1+\bm{\chi}_2)^2}\right)\qquad\qquad\qquad\qquad\qquad\qquad\nonumber\\
\cdot\int d^2\theta\frac{\bm{\theta}-\bm{\chi}_2}{\gamma^{-2}+(\bm{\theta}-\bm{\chi}_2)^2}\cos\frac{\omega T}{2}\left[\chi_2^2+2\bm{\chi}_2\cdot(\bm{\theta}-\bm{\chi}_2)\right]\nonumber\\
\nonumber\\
\simeq -\frac{\bm{\chi}_2\cdot(\bm{\chi}_1+\bm{\chi}_2)}{(\bm{\chi}_1+\bm{\chi}_2)^2}\frac{2}{\gamma\chi_1}\sin\frac{\omega T\chi_1^2}{2}K_1\left(\frac{\omega T\chi_1}{\gamma}\right)\qquad\quad\nonumber\\
-\frac{\bm{\chi}_1\cdot(\bm{\chi}_1+\bm{\chi}_2)}{(\bm{\chi}_1+\bm{\chi}_2)^2}\frac{2}{\gamma\chi_2}\sin\frac{\omega
T\chi_2^2}{2}K_1\left(\frac{\omega
T\chi_2}{\gamma}\right),\qquad\quad
\end{eqnarray}
with $K_1$ the Macdonald function \cite{Abr-Stegun}. At relatively
low $\omega$ ($\chi^{-2}\ll\omega T\ll\gamma\chi^{-1}$), when
$K_1\left(\frac{\omega T\chi}{\gamma}\right)\to\frac{\gamma}{\omega
T\chi}$, form (\ref{interm-omega-K1}) reduces to the high-$\omega$
asymptotics (\ref{asympt-sin-complex-form}) of Eq.~(\ref{FA-def}).
Thus, equations (\ref{FA-def}) and (\ref{interm-omega-K1}) can be
unified by writing
\begin{eqnarray}\label{FAFj}
r\left(\bm{\chi}_1,\bm{\chi}_2,\gamma^{-1},\omega
T\right)\qquad\qquad\qquad\qquad\qquad\nonumber\\
\simeq A_1\!\left(\frac{\omega T\chi_1^2}2,\frac{\omega T\chi_1{\chi}_2}2 e^{i\varphi_{12}}\!\right)\!F_{\perp}\!\left(\frac{\omega T\chi_1}{\gamma}\right)\nonumber\\
+A_1\!\left(\frac{\omega T\chi_2^2}2,\frac{\omega T\chi_1{\chi}_2}2
e^{i\varphi_{12}}\!\right)\!F_{\perp}\!\left(\frac{\omega
T\chi_2}{\gamma}\right)\nonumber\\
+A_2\!\left(\frac{\omega T\chi_1{\chi}_2}2
e^{i\varphi_{12}}\!\right).\qquad\qquad \qquad\qquad\,
\end{eqnarray}
Here
\begin{eqnarray}\label{A1-def}
A_1\left(z_1,z_2\right)=-\text{Ci}\left(z_1\right)\qquad\qquad\qquad\qquad\qquad\quad\nonumber\\
+\mathfrak{Re}\left\{\cos z_2\text{Ci}\left(z_1+z_2\right)+\sin
z_2\text{si}\left(z_1+z_2\right)\right\},
\end{eqnarray}
\begin{equation}\label{A2-def}
A_2\left(z\right)=-\mathfrak{Re}\left\{\cos z
\text{Ci}\left(z\right)+\sin z\text{si}\left(z\right)\right\},
\end{equation}
may be interpreted as quasiantenna form factors, and
\begin{equation}\label{Fj-def}
F_{\perp}(z)=zK_1(z),
\end{equation}
being normalized by condition $F_{\perp}(0)=1$, as the electron's
proper field form factor. In the next section, we will investigate
its origin in more detail. Term $A_2$ in (\ref{FAFj}) (stemming from
the low-$\theta$ part of interference between the external lines) at
$\omega T\chi^2\gg1$ decreases faster than $A_1$:
\begin{equation}\label{A2simz-2}
A_2(z)\underset{z\to\infty}\simeq \mathfrak{Re}\frac1{z^2}.
\end{equation}
This coincides with the transient asymptotics of the original
integral as a whole, so it appears unnecessary to endow $A_2$ with a
suppressing form factor.\footnote{With the account of
$\mathcal{O}(\gamma^{-1})$ corrections, (\ref{A2simz-2}) is actually followed by slow oscillations $-\frac{1}{\omega
T\chi^2}\sin\frac{\omega T}{2\gamma^2}$, which ensure that
$\int_0^{\infty}d\omega A_2$ exactly equals zero, as is
$\int_0^{\infty}d\omega A_1$. However, that faint contribution is
virtually invisible against $I_{\text{BH}}$ and $g$, so it seems
harmless to neglect it entirely.}

Sine factors in (\ref{interm-omega-K1}) produce oscillations similar
to those in (\ref{2I0+cos}), but are related with a different
(electron scattering angle dependent) definition of the photon
formation length:
\begin{equation}\label{resonance-condition-soft}
\sin\frac{\omega T\chi^2}{2}=\sin\frac{T}{l_{\chi}(\omega)}, \qquad
l_{\chi}(\omega)\underset{\gamma\chi\gg
1}\simeq\frac2{\omega\chi^2}.
\end{equation}
The reason why, in contrast to Eq.~(\ref{2I0+cos}), we encounter
here a sine instead of cosine dependence is that in
Eq.~(\ref{h-single-int}), cosine functions are integrated over
photon emission angles from $\chi_1^2$, $\chi_2^2$ to infinity.
Ultimately, those oscillations are damped by the exponentially
decreasing factor $F_{\perp}$, but the damping proceeds slowly,
since $F_{\perp}$ depends on $\omega$ on a scale which is
$\gamma\chi$ times harder than the arguments of $A$'s. So, there is
enough room for the spectrum to make a number of visible
oscillations.

\begin{figure}
\includegraphics{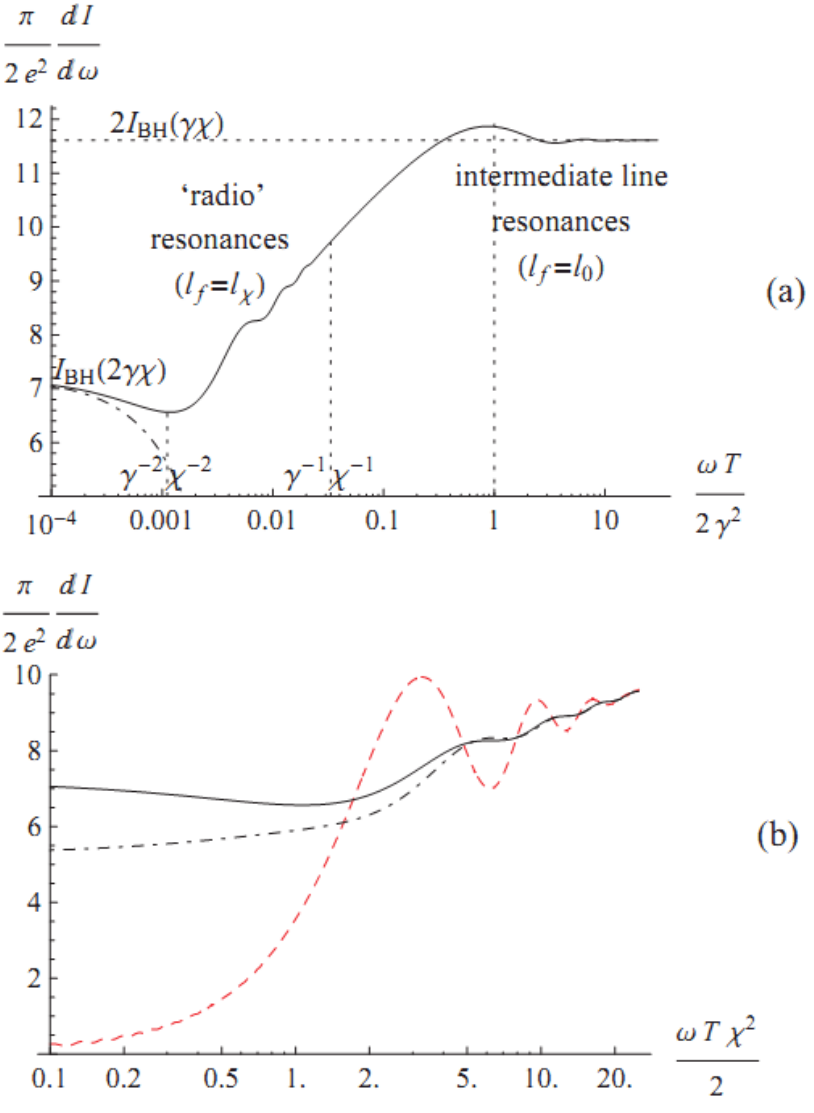}
\caption{\label{fig:Oscill} The spectrum of electromagnetic
radiation from a doubly scattered electron. (a) Full spectrum for a
case of scattering through equal angles
$\bm{\chi}_1=\bm{\chi}_2=\bm{\chi}$, $\chi=30\gamma^{-1}$ (solid
curve). Approximation (\ref{dIdomega-sum-formfactors}) is virtually
indistinguishable from this. The dot-dashed curve shows the behavior
of asymptotics (\ref{IR-asympt-NLO}). Two separate regions of
oscillatory behavior at intermediate and at high $\omega$ are
determined by different photon formation lengths. (b) Soft region of
the radiation spectrum, for a more general case
$|\bm{\chi}_1|=|\bm{\chi}_2|=30\gamma^{-1}$ and different values of
azimuth $\varphi_{12}$. Solid curve, $\varphi_{12}=0$ (as in the
upper figure). Dot-dashed, $\varphi_{12}=3\pi/4$ [evaluated by
Eq.~(\ref{dIdomega-sum-formfactors})]. Red dashed curve,
$\varphi_{12}=\pi$, corresponding to the jet overlap is evaluated
by exact representations (\ref{g-def}) and (\ref{h-def}). In the
latter case, the oscillations are anomalously large.}
\end{figure}

Contribution (\ref{FAFj}) may now be added to
(\ref{dIdomega-highomega}), and their sum
\begin{eqnarray}\label{dIdomega-sum-formfactors}
\frac{dI}{d\omega}\underset{\chi_{1,2}\gg\gamma^{-1}}\simeq \frac{dI_{\text{BH}}}{d\omega}(\gamma\chi_1)+\frac{dI_{\text{BH}}}{d\omega}(\gamma\chi_2)\qquad\qquad\qquad\qquad\quad\nonumber\\
+\frac{2e^2}{\pi}\Bigg[g\left(\frac{\omega T}{2\gamma^2}\right)+A_1\left(\frac{\omega T\chi_1^2}2,\frac{\omega T\chi_1{\chi}_2}2 e^{i\varphi_{12}}\!\right)F_{\perp}\left(\frac{\omega T\chi_1}{\gamma}\right)\,\nonumber\\
+A_1\!\left(\frac{\omega T\chi_2^2}2,\frac{\omega T\chi_1{\chi}_2}2
e^{i\varphi_{12}}\!\right)F_{\perp}\!\left(\frac{\omega
T\chi_2}{\gamma}\right)\nonumber\\
+A_2\!\left(\frac{\omega T\chi_1{\chi}_2}2
e^{i\varphi_{12}}\!\right)\!\Bigg]\qquad \qquad\qquad\qquad
\end{eqnarray}
gives a satisfactory approximation to the exact result for all
$\omega$. Term $g$ here represents a hard contribution,
whereas the residual quasiantenna terms represent the soft contribution,
which is yet regulated by the appropriate proper field formfactors
depending on $\omega$ on a scale intermediate between soft and hard.
At $\omega\to0$, logarithmic divergences of terms $g$ and $A_2$
mutually cancel, as they must, and up to terms linear in $\omega$,
the result reads
\begin{equation}\label{IR-asympt-NLO}
\frac{dI}{d\omega}\underset{\omega T\chi_{1,2}^2\ll1}\simeq
\frac{dI_{\text{BH}}}{d\omega}(\gamma|\bm{\chi}_1+\bm{\chi}_2|)-e^2\omega
T\frac{\bm{\chi}_1\cdot\bm{\chi}_2}{2}.
\end{equation}
Note that the last term here is negative when
$\bm{\chi}_1\cdot\bm{\chi}_2>0$; hence, in the low-$\omega$ domain,
the spectrum suppression can be nonmonotonous and dive below the
infrared limiting value.

A typical resulting spectrum for significant electron deflection
angles (which are let have equal values, $\bm{\chi}_1=\bm{\chi}_2$) is
shown in Fig.~\ref{fig:Oscill}(a). It displays oscillations in two
nonoverlapping regions, with visibilities $\sim1/\ln\gamma^2\chi^2$.
If $|\bm{\chi}_1|$ and $|\bm{\chi}_2|$ were unequal, according to
Eq.~(\ref{interm-omega-K1}), low-omega oscillations would involve
two periods, becoming less regular.\footnote{If furthermore we
average over an interval of $\chi_1$ and $\chi_2$ so large that
$\delta\chi_{1,2}\gtrsim\chi_{1,2}$, soft oscillations would be
washed out. That is why they were not discovered in works
\cite{Blankenbecler,BK-structured,Zakharov,Bondarenco-Shulga}.} At
lowest $\omega$, in Fig.~\ref{fig:Oscill}a there is a bump predicted
by Eq.~(\ref{IR-asympt-NLO}). In practice, a superficially similar
bumplike structure often occurs due to transition radiation on solid
target boundaries (see \cite{Klein}). However, we do not include
here any medium polarization effects, the bump being merely due to
positive correlation between electron deflection angles.

Comparing solid and dot-dashed curves in Fig.~\ref{fig:Oscill}(b),
we can see that for the case of scattering angles coinciding by
absolute value, low-$\omega$ oscillations are virtually independent
of the azimuth between the scattering planes, given that in
Eq.~(\ref{interm-omega-K1}),
$\frac{\bm{\chi}_1\cdot(\bm{\chi}_1+\bm{\chi}_2)}{(\bm{\chi}_1+\bm{\chi}_2)^2}=\frac{\bm{\chi}_2\cdot(\bm{\chi}_1+\bm{\chi}_2)}{(\bm{\chi}_1+\bm{\chi}_2)^2}=\frac12$.
Anomalously large oscillations emerge, however, at
$\bm{\chi}_2\to-\bm{\chi}_1$ (dashed curve). That corresponds to an
overlap of radiation cones aligned with initial and final electron
lines, and leads to breakdown of formula (\ref{interm-omega-K1}).
Such a case is exceptional, and generally will be beyond the scope
of the present paper.

The causal connection between directions of electron motion and that
of photon emission means that together they form a jet. More
precisely, in this process there are two categories of photons:
int\emph{ra}jet (inside  a jet) and int\emph{er}jet (between the
jets), as is evident from Fig.~\ref{fig:Ang}. In quantum electrodynamics, in
interpretations of radiative corrections integrated over $\omega$
and all components of $\bm{k}$, intrajet photons are generally
called collinear, whereas low-frequency photons which do not have
collinear properties (in our case -- interjet, although they may
incidentally propagate along one of the jet directions, as well) are
called soft \cite{collinear-soft-photons,Dokshitzer}.
In their terms, internal line resonances in the hard spectral domain
in Fig.~\ref{fig:Oscill}(a) are due to ``collinear-collinear"
radiation interference (interference between photons generated by
the electron in the intermediate state and emitted close to its
velocity), whereas ``radio" resonances in the soft domain are
``soft-collinear" interference (when only one of the interfering
photons is closely aligned with the initial or final electron line).

The notion of jets also helps elucidate why photon formation length
(\ref{resonance-condition-soft}) results from generic
Eq.~(\ref{coh-length-theta}) by \emph{exactly} substituting
$\theta\to\chi$: The emission angle for interfering photons is
counted off from the direction of one of the electron lines
(internal) to the direction of another (external) electron line,
along which such photons are actually emitted, and the indeterminacy of the emission angle $\sim\gamma^{-1}$ is much smaller than its mean value $\chi$. This, though, does
not completely specify the process geometry in position space yet.
There also remains an issue why the proper field form factors, which
are asymptotically exponential, depend on the absolute value of the
deflection angle. Finally, our assumption that $l_0$ and $l_{\chi}$
are the photon formation lengths was actually not strictly proven
within the approach of the present section. It thus deserves
additional space-time considerations. In particular, one can
anticipate the factorization property for the low-$\omega$ part also
to be backed by some spatially causal reasons.

\section{Impact parameter representation}\label{subsec:imp-par}

In this section, we will explore properties of transverse spatial
variables for emitted photons. They must be Fourier-dual to the
photon transverse wave vector, and actually be in the spirit of ray
optics. If conditions of ray optics do apply, impact parameters\footnote{Here we actually deal with photon emission rather than impact, but to stress the analogy with the equivalent photon method, we adopt the same terminology.}
should assume rather well-defined values characterizing preferable
light rays.

Some complications emerge in this regard, however, since
Eq.~(\ref{dIdomega-through-amp}) involves not the local
electromagnetic field, but the radiation emission amplitude. Besides
that, the electromagnetic field is physically coupled to the
electron, which arrives from and moves off to infinity. Nonetheless, well
defined should be the notion of impact parameter of an
electromagnetic wave with respect to the initial or to the final
electron line. At that, since directions of electron motion along
those lines differ, there will arise simultaneously two instead of
one species of the impact parameter (in contrast to the Glauber description
of short-wave scattering on finite obstacles, where a single definition for impact parameters is sufficient).

A formulation of the impact parameter view for the case of electron
double scattering can be attained as follows.
In representation (\ref{dIdomega=intd2theta}),
for each of the algebraic terms in the brackets, apply transformation
\[
\frac{\bm{\theta}}{\gamma^{-2}+\theta^2}=\frac{i}{2\pi}\int d^2\xi
e^{i\bm{\theta}\cdot\bm{\xi}}\frac{\partial}{\partial
\bm{\xi}}K_0\left(\frac{\xi}{\gamma}\right),
\]
[obtained from well known identity
$\frac{1}{\gamma^{-2}+\theta^2}=\frac{1}{2\pi}\int d^2\xi
e^{i\bm{\theta}\cdot\bm{\xi}}K_0\left(\frac{\xi}{\gamma}\right)$ by integrating by parts].
That leads to representation \cite{Bondarenco-Shulga}, which for our
present purposes more conveniently casts as
\begin{eqnarray}\label{imp-par}
\frac{dI}{d\omega}=\frac{dI_{\text{BH}}}{d\omega}(\gamma\chi_1)+\frac{dI_{\text{BH}}}{d\omega}(\gamma\chi_2)\qquad\qquad\qquad\qquad\qquad\nonumber\\
-\frac{e^2}{\pi^3\omega T}\iint d^2 \xi_1 d^2 \xi_2
\frac{\partial}{\partial
\bm{\xi}_1}K_0\left(\frac{\xi_1}{\gamma}\right)
\cdot\frac{\partial}{\partial
\bm{\xi}_2}K_0\left(\frac{\xi_2}{\gamma}\right)
 \nonumber\\
\times \mathfrak{Im}\left(1-e^{-i\bm{\chi}_1\cdot \bm{\xi}_1}
\right)\left(1-e^{-i\bm{\chi}_2\cdot \bm{\xi}_2} \right)
e^{-i\frac{\omega
T}{2\gamma^{2}}+i\frac{(\bm{\xi}_1-\bm{\xi}_2)^2}{2\omega T}}.
\end{eqnarray}
The impact parameter here is represented by $\bm{\xi}/\omega$ rather
than $\bm{\xi}$ alone (the latter is dimensionless). Specifically,
$\bm{\xi}_1/\omega$ is the impact parameter with respect to the
first scattering vertex, and $\bm{\xi}_2/\omega$ is that with
respect to the second vertex. We will see shortly that this approach is
largely similar to the equivalent photon one \cite{Jackson}, with a
proviso that from the outset we deal with strictly real photons, and
do not restrict ourselves to dipole approximation.

Examining structure (\ref{imp-par}), it is evident that at
$\chi\gg\gamma^{-1}$, terms $e^{-i\bm{\chi}_1\cdot \bm{\xi}_1}$ and
$e^{-i\bm{\chi}_2\cdot \bm{\xi}_2}$ are rapidly oscillating. Thus,
the main contribution to the integral is brought by the
$\chi$-independent term
\begin{eqnarray}\label{Ci-si-imp-par}
g\!\left(\frac{\omega T}{2\gamma^2}\right)\!=\frac{1}{2\pi^2\omega
T}\!\iint \! d^2 \xi_1 d^2 \xi_2 \frac{\partial}{\partial
\bm{\xi}_1}K_0\!\left(\frac{\xi_1}{\gamma}\right)
\!\cdot\!\frac{\partial}{\partial
\bm{\xi}_2}K_0\!\left(\frac{\xi_2}{\gamma}\right)\nonumber\\
\times\sin\left[\frac{\omega
T}{2\gamma^{2}}-\frac{(\bm{\xi}_1-\bm{\xi}_2)^2}{2\omega
T}\right],\qquad\quad
\end{eqnarray}
which can be shown (e.g., by returning to the emission angle
representation) to coincide with (\ref{g-def}). The high-$\omega$
asymptotics of Eq.~(\ref{Ci-si-imp-par}) can be derived by noting
that therein typical contributing $\xi_1$, $\xi_2$ are small. It is,
however, impossible to entirely neglect term
$\frac{(\bm{\xi}_1-\bm{\xi}_2)^2}{2\omega T}$ in the argument of the
sine, because then the integrals over $\bm{\xi}_1$ and
$\bm{\xi}_2$ would vanish. Expanding through the next-to-leading
order
\begin{eqnarray}\label{small-impact-param}
\sin\left[\frac{\omega
T}{2\gamma^{2}}-\frac{(\bm{\xi}_1-\bm{\xi}_2)^2}{2\omega
T}\right]\simeq \sin\left(\frac{\omega
T}{2\gamma^{2}}-\frac{\bm{\xi}_1^2+\bm{\xi}_2^2}{2\omega T}\right)\nonumber\\
+\frac{\bm{\xi}_1\cdot\bm{\xi}_2}{\omega T}\cos\frac{\omega
T}{2\gamma^{2}},
\end{eqnarray}
and inserting this to (\ref{Ci-si-imp-par}) reproduces
Eq.~(\ref{2I0+cos}).

In contrast, at low $\omega$, we know from Sec.~\ref{subsec:2.1}
that the rest of the interference terms become important, as well,
but here it is justified to set $\gamma^{-1}\to0$. From the
standpoint of representation (\ref{imp-par}), that owes to the
smallness of contributing $\xi_1$, $\xi_2$. Integration in
\begin{eqnarray*}
r\left(\bm{\chi}_1,\bm{\chi}_2,0,\omega T\right)=\frac1{2\pi^2\omega T}\iint d^2\xi_1 d^2\xi_2\frac{\bm{\xi}_1}{\xi_1^2}\cdot\frac{\bm{\xi}_2}{\xi_2^2}\qquad\qquad\nonumber\\
\times\mathfrak{Im}\left(e^{-i\bm{\chi}_1\cdot
\bm{\xi}_1}+e^{-i\bm{\chi}_2\cdot \bm{\xi}_2}-e^{-i\bm{\chi}_1\cdot
\bm{\xi}_1-i\bm{\chi}_2\cdot
\bm{\xi}_2}\right)e^{i\frac{(\bm{\xi}_1-\bm{\xi}_2)^2}{2\omega T}}
\end{eqnarray*}
with the use of same conformal properties (\ref{conf-int-1}) and
(\ref{conf-int-2}) gives back the same integral sine and cosine
representation (\ref{FA-def}).

Our main interest, however, lies in the case $\omega T\chi^2\gg1$,
since impact parameters then assume rather sharp values, and suggest a direct physical interpretation. We thus examine integral
(\ref{imp-par}) under condition $\chi^{-2}\ll\omega
T\ll\chi^{-1}\gamma$ more closely.

To begin with, in one of the $\chi$-dependent terms of
(\ref{imp-par}),
\begin{eqnarray*}
\frac{1}{2\pi^2\omega T}\mathfrak{Im}\iint d^2 \xi_1 d^2 \xi_2
\frac{\partial}{\partial
\bm{\xi}_1}K_0\left(\frac{\xi_1}{\gamma}\right)
\cdot\frac{\partial}{\partial
\bm{\xi}_2}K_0\left(\frac{\xi_2}{\gamma}\right)\nonumber\\
\times e^{-i\bm{\chi}_1\cdot \bm{\xi}_1+\frac{i}{2\omega
T}(\bm{\xi}_1-\bm{\xi}_2)^2},
\end{eqnarray*}
the dominant contribution stems from small $\xi_2$, allowing it to
be approximated as
\begin{eqnarray}\label{1st-imp-par-int}
-\frac{1}{2\pi^2\omega T}\mathfrak{Im}\int d^2 \xi_1
\frac{\partial}{\partial
\bm{\xi}_1}K_0\left(\frac{\xi_1}{\gamma}\right)e^{-i\bm{\chi}_1\cdot
\bm{\xi}_1+\frac{i}{2\omega T}\xi_1^2} \nonumber\\
\cdot\int d^2 \xi_2\frac{\bm{\xi}_2}{\xi_2^2}e^{-\frac{i}{\omega
T}\bm{\xi}_1\cdot\bm{\xi}_2}.\qquad\qquad\quad
\end{eqnarray}
Here the integral over $\bm{\xi}_2$ equals $\int d^2
\xi_2\frac{\bm{\xi}_2}{\xi_2^2}e^{-\frac{i}{\omega
T}\bm{\xi}_1\cdot\bm{\xi}_2}=\frac{2\pi\omega
T}{i}\frac{\bm{\xi}_1}{\xi_1^2}$, and that over $\bm{\xi}_1$ engages
a rapidly oscillating exponential $e^{-i\bm{\chi}_1\cdot
\bm{\xi}_1+\frac{i}{2\omega T}\xi_1^2}=e^{\frac{i}{2\omega
T}(\bm{\xi}_1-\omega T\bm{\chi}_1)^2-\frac{i}2\omega T\chi_1^2}$,
which has a stationary phase point at $\bm{\xi}_1=\omega
T\bm{\chi}_1$. That effectively fixes $\bm{\xi}_1$ in other factors
at this value:
\[
\frac{\partial}{\partial
\bm{\xi}_1}K_0\left(\frac{\xi_1}{\gamma}\right)\cdot\frac{\bm{\xi}_1}{\xi_1^2}\to-\frac{1}{\gamma\omega
T\chi_1}K_1\left(\frac{\omega T\chi_1}{\gamma}\right).
\]
The result of integration in (\ref{1st-imp-par-int}) then
equals
\begin{equation}\label{sin-K1-1}
-\frac2{\gamma\chi_1}\sin\frac{\omega
T\chi_1^2}2K_1\left(\frac{\omega T\chi_1}{\gamma}\right).
\end{equation}

Similarly, the integral containing
$e^{-i\bm{\chi}_2\cdot\bm{\xi}_2}$ reduces to
\begin{eqnarray}\label{sin-K1-2}
\frac{1}{2\pi^2\omega T}\mathfrak{Im}\iint d^2 \xi_1 d^2 \xi_2
\frac{\partial}{\partial
\bm{\xi}_1}K_0\left(\frac{\xi_1}{\gamma}\right)
\cdot\frac{\partial}{\partial
\bm{\xi}_2}K_0\left(\frac{\xi_2}{\gamma}\right)\nonumber\\
\times e^{-i\bm{\chi}_2\cdot \bm{\xi}_2+\frac{i}{2\omega
T}(\bm{\xi}_1-\bm{\xi}_2)^2}\nonumber\\
\simeq -\frac2{\gamma\chi_2}\sin\frac{\omega
T\chi_2^2}2K_1\left(\frac{\omega
T\chi_2}{\gamma}\right).\qquad\qquad\qquad\qquad
\end{eqnarray}

Finally, the integral containing
$e^{-i\bm{\chi}_1\cdot\bm{\xi}_1-i\bm{\chi}_2\cdot\bm{\xi}_2}$
receives two dominant contributions, in one of which $\bm{\xi}_1$ is
small while $\bm{\xi}_2$ is finite, and in the other one
$\bm{\xi}_2$ is small while $\bm{\xi}_1$ is finite:
\begin{eqnarray*}\label{}
-\frac{1}{2\pi^2\omega T}\mathfrak{Im}\iint d^2 \xi_1 d^2 \xi_2
\frac{\partial}{\partial
\bm{\xi}_1}K_0\left(\frac{\xi_1}{\gamma}\right)
\cdot\frac{\partial}{\partial
\bm{\xi}_2}K_0\left(\frac{\xi_2}{\gamma}\right)\nonumber\\
\times e^{-i\bm{\chi}_1\cdot \bm{\xi}_1-i\bm{\chi}_2\cdot
\bm{\xi}_2+\frac{i}{2\omega
T}(\bm{\xi}_1-\bm{\xi}_2)^2}\nonumber\\
\simeq \frac{1}{2\pi^2\omega T}\mathfrak{Im}\int d^2 \xi_2
e^{-i\bm{\chi}_2\cdot \bm{\xi}_2+\frac{i}{2\omega T}\xi_2^2}
\frac{\partial}{\partial \bm{\xi}_2}K_0\left(\frac{\xi_2}{\gamma}\right)\nonumber\\
\cdot \int d^2 \xi_1\frac{\bm{\xi}_1}{\xi_1^2}
e^{-i\left(\bm{\chi}_1+\frac{\bm{\xi}_2}{\omega T}\right)\cdot \bm{\xi}_1}\nonumber\\
+\frac{1}{2\pi^2\omega T}\mathfrak{Im}\int d^2 \xi_1
e^{-i\bm{\chi}_1\cdot \bm{\xi}_1+\frac{i}{2\omega T}\xi_1^2}
\frac{\partial}{\partial \bm{\xi}_1}K_0\left(\frac{\xi_1}{\gamma}\right)\nonumber\\
\cdot\int d^2 \xi_2 \frac{\bm{\xi}_2}{\xi_2^2}
e^{-i\left(\bm{\chi}_2+\frac{\bm{\xi}_1}{\omega T}\right)\cdot
\bm{\xi}_2}.
\end{eqnarray*}
Those integrals can be evaluated in exactly the same way as
above, giving
\begin{eqnarray*}\label{}
-\frac{1}{2\pi^2\omega T}\mathfrak{Im}\iint d^2 \xi_1 d^2 \xi_2
\frac{\partial}{\partial
\bm{\xi}_1}K_0\left(\frac{\xi_1}{\gamma}\right)
\cdot\frac{\partial}{\partial
\bm{\xi}_2}K_0\left(\frac{\xi_2}{\gamma}\right)\nonumber\\
\times e^{-i\bm{\chi}_1\cdot \bm{\xi}_1-i\bm{\chi}_2\cdot
\bm{\xi}_2+\frac{i}{2\omega
T}(\bm{\xi}_1-\bm{\xi}_2)^2}\nonumber\\
\simeq \frac{2}{\gamma
\chi_2}\frac{\bm{\chi}_2\cdot\left(\bm{\chi}_1+\bm{\chi}_2\right)}{\left(\bm{\chi}_1+\bm{\chi}_2\right)^2}\sin\frac{\omega
T\chi_2^2}2
K_1\left(\frac{\omega T\chi_2}{\gamma}\right)\nonumber\\
+\frac{2}{\gamma
\chi_1}\frac{\bm{\chi}_1\cdot\left(\bm{\chi}_1+\bm{\chi}_2\right)}{\left(\bm{\chi}_1+\bm{\chi}_2\right)^2}\sin\frac{\omega
T\chi_1^2}2 K_1\left(\frac{\omega T\chi_1}{\gamma}\right).
\end{eqnarray*}
Combined with (\ref{sin-K1-1}) and (\ref{sin-K1-2}), it leads to result
(\ref{interm-omega-K1}).

\begin{figure}
\includegraphics{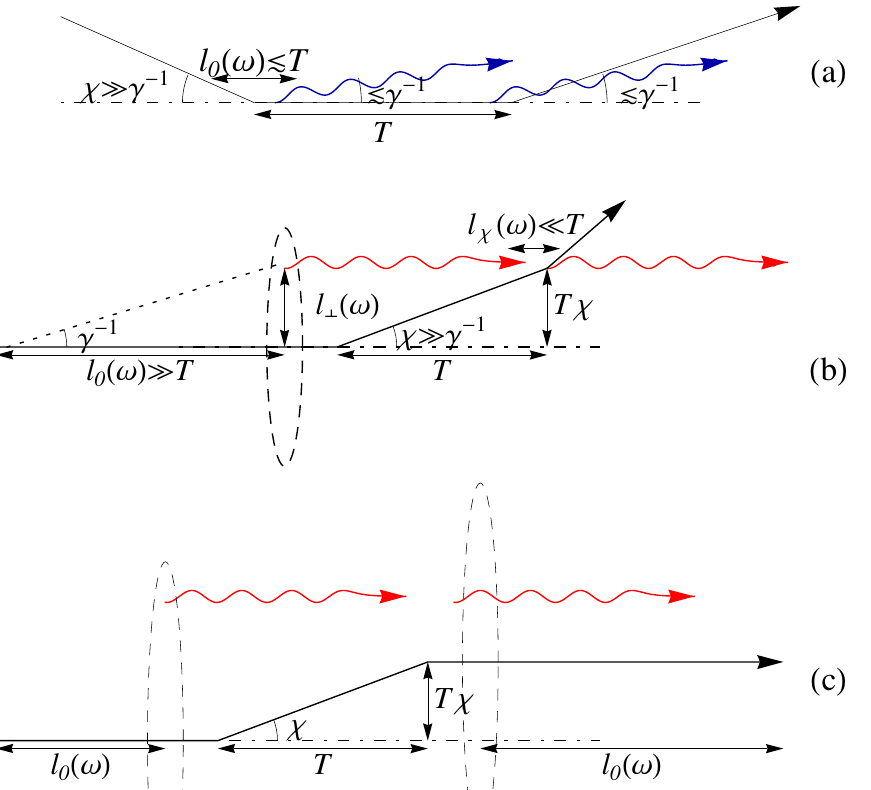}
 \caption{\label{fig:Diagr} (a) High-$\omega$ interference diagram. (b) Low-$\omega$ interference diagram, in the absence of jet overlap.
In this case, there is also a cross diagram, in which noncollinear
photons are emitted from the first scattering vertex, and collinear
ones from the final electron line. The condition of interference
between collinear and noncollinear photons, besides the
coincidence of emission directions, is the equality of impact
parameters: $l_{\perp}(\omega)=T\chi$. (c) Low-$\omega$
interference diagram in the case of jet overlap
($\bm{\chi}_2=-\bm{\chi}_1$). For interfering collinear photons in
this case, only the difference of impact parameters is fixed.}
\end{figure}

The derivation offered above elucidates the geometrical origin of
proper field form factor $F_{\perp}$: It corresponds to the impact
parameter distribution amplitude for one of the intra jet photons,
at a fixed impact parameter $\Delta
\bm{b}=\bm{\xi}_1/\omega=T\bm{\chi}_1$ determined by the difference
of impact parameters between the vertices, since emission of an
interjet photon is completed in a relatively small spatial domain. The
corresponding form factors in Eq.~(\ref{dIdomega-sum-formfactors}),
\[
F_{\perp}\left(\frac{\omega
T\chi}{\gamma}\right)=F_{\perp}\left(\frac{\Delta
b(T,\chi)}{l_{\perp}(\gamma,\omega)}\right),
\]
depend on the ratio of the aforementioned $\Delta b$ and
\begin{equation}\label{lperp-def}
l_{\perp}(\omega)=\frac{\gamma}{\omega},
\end{equation}
which thus plays the role of a ``transverse coherence length". One
actually recognizes in (\ref{lperp-def}) nothing but the typical
transverse scale for the electric field of an ultrarelativistic
particle, well known within, e.g., the equivalent photon approach
\cite{Jackson}. In capacity of a coherence length in our problem, it
does not give rise to a new type of oscillations, but describes
damping of an old one.

The analogy with the equivalent photon approach is strengthened by
observing that Fourier expansion of the transverse component of the
electric or magnetic field of ultrarelativistic electron
$E_{\perp}(b,z,t)=\frac{Ze\gamma
b}{\left[b^2+\gamma^2(z-vt)^2\right]^{3/2}}$ gives
\[
\int_{-\infty}^{\infty}dt e^{i\omega
t}E_{\perp}(b,0,t)=\frac{2Ze}{vb}F_{\perp}\left(\frac{\omega b}{v\gamma}\right),
\]
where $v\to1$ in the ultrarelativistic case, and form factor
$F_{\perp}$, absorbing all the $\omega$-dependence, coinciding with
(\ref{Fj-def}).

It may be instructive to compare virtues of the formalism of the
present section with that in Sec.~\ref{subsec:2.1}. In the hard
spectral domain, the interference of radiation is described well
enough by the photon emission angle representation of
Sec.~\ref{subsec:2.1}, demonstrating that photons participating in the
interference emerge under fairly well-defined angles close to the
intermediate electron direction of motion [see
Fig.~\ref{fig:Diagr}(a)]. The range of contributing angles shrinks
reciprocally with the increase of $\omega$:
\begin{equation}\label{delta-theta2}
(\delta \theta)^2\sim\frac{2}{\omega T},
\end{equation}
producing the power-law falloff factor in the spectral oscillations.
In the soft domain, however, a clearer physical picture is offered
by the impact parameter representation, revealing that in addition
to emission of the interfering photons parallel to the initial or
final electron line [as is already clear in the emission angle
representation of Sec.~\ref{subsec:2.1}, particularly
Fig.~\ref{fig:Ang}(b)], in the configuration space they must yet
nearly belong to a ray going parallel to the external electron line
at a distance such that it passes through the opposite vertex (see
Fig.~\ref{fig:Diagr}b). At that, the fraction of such photons, quantified by the spread of the contributing impact parameters in integral (\ref{imp-par}), $\delta \left(\xi/\omega\right)^2\sim T/\omega$, appears to be significant  compared to $(T\chi)^2$ if $\omega T\sim\chi^{-2}$.
With the increase of $\omega$, this fraction decreases, because of
the exponential decrease of the intrajet photon wave function at
large impact parameters.

The impact parameter view also gives better understanding of the
condition of applicability of classical electrodynamics for the present process: The
requirement of negligibility of photon recoil,
$\omega\ll{E}/{\hbar}$, in combination with our estimate
$\omega\sim\frac1{T\chi^2}$ for typical photon frequencies in the
soft interference domain expresses as
\[
p_{\perp}\Delta b\gg\hbar,
\]
where $p_{\perp}=E\chi$ is the electron transverse momentum, while
$\Delta b=T\chi$, as before. Hence, the semiclassical tractability
of soft photon emission in the given process is equivalent to the
semiclassicality of \emph{transverse motion of the electron} within
the intermediate trajectory segment.

\section{Time evolution}\label{sec:double-time}

To corroborate our conjecture that $l_0$ and $l_{\chi}$ are the true
formation lengths for interfering photons in the corresponding
spectral domains, it is necessary to provide also some longitudinal
coordinate considerations. It is difficult to simultaneously handle
all three spatial coordinates and the photon emission frequency, so
we will restrict ourselves in this section to a simplified treatment
only in terms of the photon emission times, which are manifestly
present in formula (\ref{dIdomega-through-amp}).

An emission time representation for the radiation spectrum derives
from (\ref{dIdomega-through-amp}) by performing prior integration
over the radiation angles, which is manageable for a generic electron
trajectory $\bm{r}(t)$. That relinquishes the issue of the photon
formation length dependence on the emission angles, but instead
introduces its direct dependence on the electron deflection angles.
Moreover, since we now encounter a double time integral, coherence
lengths duplicate, and there may also occur a correlation between
the emission times. We will expound the corresponding procedure as
briefly as possible.

In the photon emission spectrum, two-time correlation on the electron trajectory, in effect, is
mediated by the photon propagator (see the Appendix), which depends
on the electromagnetic field gauge. The simplest for use is Feynman
gauge, in which the angle-integral radiation spectrum reads:
\begin{eqnarray}\label{t2tau-subtr}
\frac{dI}{d\omega}
=\omega\frac{e^2}{\pi}\int_0^{\infty}\frac{d\tau}{\tau}\int_{-\infty}^{\infty}\!dt_2\Bigg(\!\! \left\{\gamma^{-2}\!+\!\frac12\!\left[\bm{v}(t_2)- \bm{v}(t_2-\tau)\right]^2\right\}\nonumber\\
\qquad\qquad\qquad\times \sin \omega \left[\tau-\left|\bm{r}(t_2)-\bm{r}(t_2-\tau)\right|\right]\nonumber\\
\qquad\qquad\qquad\qquad\qquad-\gamma^{-2} \sin\omega
(1-v)\tau\Bigg),\qquad\quad
\end{eqnarray}
with $\tau=t_2-t_1$. The coefficient in the argument of the last
sine here is chosen based on convenience reasons, in order to make
the integrand vanish for a uniform and rectilinear charge motion.
For $|\bm{v}(t)|=\text{const}$, this term is independent of the
particle trajectory detail.

Representation (\ref{t2tau-subtr}) is commonly used in practice, so
we will embark on it, too. Although it involves a different gauge compared to preceding sections, key elements, contained in the phase of the sine, are gauge independent.

Given the presence of two integration times, for analysis of the
coherence, the integral should be reduced to a single one with an
oscillatory integrand, leading contributions from which may
formally be related with coherence properties. In capacity of such a
variable in the present case one can take a time ratio, which,
just like photon emission angles, is related with the process
geometry. That is tantamount to dispensing with kinematic definition
(\ref{coh-length-theta}), and dealing with a more dynamical one.

For simplicity, in this section we will confine ourselves to
calculations for symmetric and coplanar electron scattering, when
\[
\bm{\chi}_1=\bm{\chi}_2.
\]
That will suffice for demonstration of relevance of coherence length
notions, and will also expose similarities with other
problems in which the electron motion is planar, e.g., synchrotron
radiation in a finite magnet.

\begin{figure}
\includegraphics{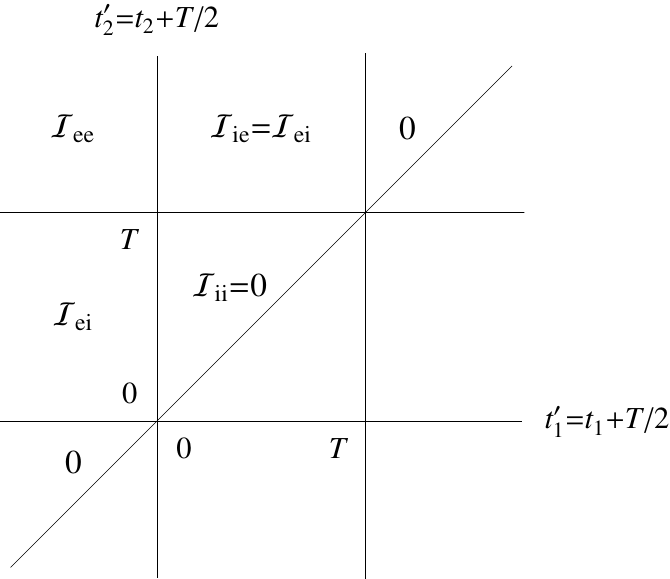}
 \caption{\label{fig:I-regions} Domains of continuity of the integrand of Eq.~(\ref{t2tau-subtr})
 in the double time plane [areas of definition of integrands of partial integrals (\ref{Iee}) and (\ref{Iei})].}
\end{figure}

Since in our case the electron trajectory is rectilinear in each of
the three intervals separated by the two scattering points, in
integral (\ref{t2tau-subtr}) nonzero are only mutual interference
terms between those intervals. Moreover, owing to the symmetry of
the trajectory with respect to its middle point, interference
integral of the inner part with the initial part is the same as that
with the final part, $\mathcal{I}_{ie}=\mathcal{I}_{ei}$ (see
Fig.~\ref{fig:I-regions}). Thus, the spectrum is comprised merely by
two essentially different contributions:
\begin{equation}\label{sum-I}
\frac{dI}{d\omega}=\frac{2e^2}{\pi}\left(\mathcal{I}_{ee}+
2\mathcal{I}_{ei}\right).
\end{equation}
In the ultrarelativistic approximation,
\begin{eqnarray}\label{Iee}
\mathcal{I}_{ee}=\frac{\omega}{2\gamma^2} \int_{T/2}^{\infty}dt_2 \int_{t_2+T/2}^{\infty}\frac{d\tau}{\tau}\Bigg\{ \left(1+2\gamma^2\chi^2\right)\qquad\qquad\nonumber\\
\times \sin \omega \left[\frac{\tau}{2\gamma^2}+\frac{\chi^2}2\left(4t_2\!\left(1-\frac{t_2}{\tau}\right)-T\right)\right]-\sin\frac{\omega\tau}{2\gamma^2}\Bigg\},
\end{eqnarray}
with $t_2$ being counted off from the midpoint of the intermediate time
interval, and
\begin{eqnarray}\label{Iei}
\mathcal{I}_{ei}=\frac{\omega}{2\gamma^2} \int_{0}^{T}dt'_2 \int_{t'_2}^{\infty}\frac{d\tau}{\tau}\Bigg\{ \left(1+\gamma^2\chi^2/2\right)\qquad\qquad\nonumber\\
\times \sin \omega \left[\frac{\tau}{2\gamma^2}+\frac{\chi^2}2 t'_2\left(1-\frac{t'_2}{\tau}\right)\right]-\sin\frac{\omega\tau}{2\gamma^2}\Bigg\}
\end{eqnarray}
with $t'_2=t_2+T/2$ counted off from the point of first scattering.
Terms non-linear in times in the arguments of sine functions
originate from transverse coordinate difference, as we expand
\begin{eqnarray}\label{vtau-|r2-r1|}
v\tau-|\bm{r}(t_2)-\bm{r}(t_2-\tau)|\simeq
\frac{1}{2v\tau}\left\{v^2\tau^2-\left[\int_{t_2-\tau}^{t_2}dt \bm{v}(t)\right]^2\right\}\nonumber\\
\simeq
\frac{1}{2v}\left\{\int_{t_2-\tau}^{t_2}dt\bm{v}_{\perp}^2(t)-\frac1{\tau}\left[\int_{t_2-\tau}^{t_2}dt
\bm{v}_{\perp}(t)\right]^2\right\}.\quad
\end{eqnarray}
Note that the latter expression in terms of transverse velocity components,
through which the longitudinal component expresses, as well, is
invariant under small rotations of the Cartesian frame.

The advantage of choosing the origin for variables $t_2$ and $t'_2$
in Eqs.~(\ref{Iee}) and (\ref{Iei}) in different points (points of
crossing of the corresponding rectilinear segments) consists in
rendering the nonlinear terms in the phase scale
invariant.\footnote{If the electron scattering angles are not
coplanar, one has to perform a complex shift of the variable $t_2$.
Thereby, complex integrals over $w$ reduce to a form analogous to
Eq.~(\ref{FA-def}).} That enables exact integration over one of the
time variables, by introducing ratio $w=2t_2/\tau$ instead of $t_2$
for $\mathcal{I}_{ee}$, and $w=t'_2/\tau$ instead of $t'_2$ for
$\mathcal{I}_{ei}$. Integrations over $\tau$ then reduce to that of
a sine with a linear argument, yielding:\footnote{In (\ref{Iee2}) we
exploited the symmetry of the integrand, by virtue of which
$\int_1^2dw\ldots=\int_0^1dw\ldots$.\label{footnote-timesymm}}
\begin{subequations}\label{eq37}
\begin{eqnarray}
\mathcal{I}_{ee}=\frac{\omega}{4\gamma^2}\int_0^2 dw\int_{\max\{\frac{T}{w}, \frac{T}{2-w}\}}^{\infty}d\tau\Bigg\{(1+2\gamma^2\chi^2)\qquad\qquad\quad\nonumber\\
\times\sin\omega\left[\frac{\tau}{2\gamma^2}+\frac{\chi^2}{2}[\tau
w(2-w)-T]\right]
-\sin\frac{\omega\tau}{2\gamma^2}\Bigg\}\nonumber\\
\label{Iee-intdwdtau}\\
= \int_{0}^{1}dw \Bigg\{\frac{1+2\gamma^2\chi^2}{1+\gamma^2\chi^2 w(2-w)} \cos\!\frac{\omega T}{2\gamma^2} \left[\frac1w +\gamma^2\chi^2 \left(1-w\right)\right]\nonumber\\
-\cos\frac{\omega T}{2\gamma^2 w}\Bigg\}\qquad\label{Iee2}
\end{eqnarray}
\end{subequations}
and
\begin{subequations}\label{Iei-intdwdtau}
\begin{eqnarray}
\mathcal{I}_{ei}=\frac{\omega}{2\gamma^2}\int_0^1dw\int_0^{T/w}d\tau\Bigg\{(1+\gamma^2\chi^2/2)\qquad\qquad\qquad\nonumber\\
\times\sin\frac{\omega\tau}{2\gamma^2}\left[1+\gamma^2\chi^2w(1-w)\right]-\sin\frac{\omega\tau}{2\gamma^2}\Bigg\}\qquad\label{Iei-intdwdtau}\\
=\int_{0}^{1}dw \Bigg\{\cos\frac{\omega
T}{2\gamma^2 w}-1\qquad\qquad\qquad\qquad\qquad\qquad\qquad\nonumber\\
+\frac{1+\gamma^2\chi^2/2}{1+\gamma^2\chi^2 w(1-w)}\left(1- \cos
\frac{\omega T}{2\gamma^2} \left[\frac1w +\gamma^2\chi^2
\left(1-w\right)\right]\!\right)\!\Bigg\}.\nonumber\\
\label{Iei2}
\end{eqnarray}
\end{subequations}

Integrals (\ref{Iee2}) and (\ref{Iei2}) span the same integration
interval, and involve identical cosine factors, so they may
reasonably be combined. Then, there arise significant cancellations
between the prefactors, which can be explicated by splitting
algebraic factors into simple fractions:
\begin{eqnarray}\label{dIdomega=intdw}
\frac{dI}{d\omega} =2\frac{dI_{\text{BH}}}{d\omega}(\gamma\chi)
+\frac{2e^2}{\pi} \int_{0}^{1}dw \Bigg\{\cos\frac{\omega T}{2\gamma^2 w}\qquad\qquad\qquad\nonumber\\
+\Bigg(\frac1{w+\frac1{2\gamma^2\chi^2}}-\frac1{w+\frac1{\gamma^2\chi^2}}
+\frac1{2-w}-\frac1{1+\frac1{\gamma^2\chi^2}-w}\Bigg)\nonumber\\
\times\cos \frac{\omega T}{2\gamma^2} \left[\frac1w +\gamma^2\chi^2
\left(1-w\right)\right] \Bigg\}.\qquad
\end{eqnarray}
Here the noninterference contribution\footnote{Structure
(\ref{I0-w}) emerges also when evaluating integral (\ref{I0}) by
Feynman parametrization. Relationships of Feynman parameters with
time variables were formerly found in quantum field theory
\cite{Coleman-Norton}.}
\begin{equation}\label{I0-w}
\frac{dI_{\text{BH}}}{d\omega}(\gamma\chi)=\frac{2e^2}{\pi}\left(\int_{0}^{1}dw\frac{1+\gamma^2\chi^2/2}{1+\gamma^2\chi^2w(1-w)}-1\right)
\end{equation}
actually coincides with (\ref{I0}).

The rest is  straightforward. From the remaining integrals, we
assess typical $w$ (either from the cosine arguments, depending on
$\omega$, or from algebraic factors), but ultimately, we need estimates for contributing times, so, to this end, we return to double
integrals (\ref{Iee-intdwdtau}) and (\ref{Iei-intdwdtau}). Note at
once that for given $w$ and $\omega$, typical $\tau$ are determined
by the slope of the $\tau$-dependence of the phase, and by
integration limits. The contribution from the end point
\begin{subequations}\label{tau,delta-tau-through-w}
\begin{equation}\label{tau-sim-Tw}
\tau\approx\frac{T}{w}
\end{equation}
in Eqs. (\ref{eq37}), (\ref{Iei-intdwdtau}) has the spread
\begin{equation}\label{delta-tau-through-w}
\delta\tau\sim\frac{2\gamma^2}{\omega\left[1+\gamma^2\chi^2w(1-w)\right]},
\end{equation}
\end{subequations}
provided $\delta\tau\lesssim\tau$, which holds for sufficiently
large $\omega$, or sufficiently small $w$. For the contribution from
the end point $\tau=0$, typical $\tau$ and $\delta\tau$ are of the
same order:
\begin{equation}\label{tau,deltatau}
\tau,\delta\tau\sim\frac{2\gamma^2}{\omega\left[1+\gamma^2\chi^2w(1-w)\right]}.
\end{equation}

We will conduct this analysis up to the full spectral
decomposition.

\subsection{Bethe-Heitler contribution}

\begin{figure}
\includegraphics{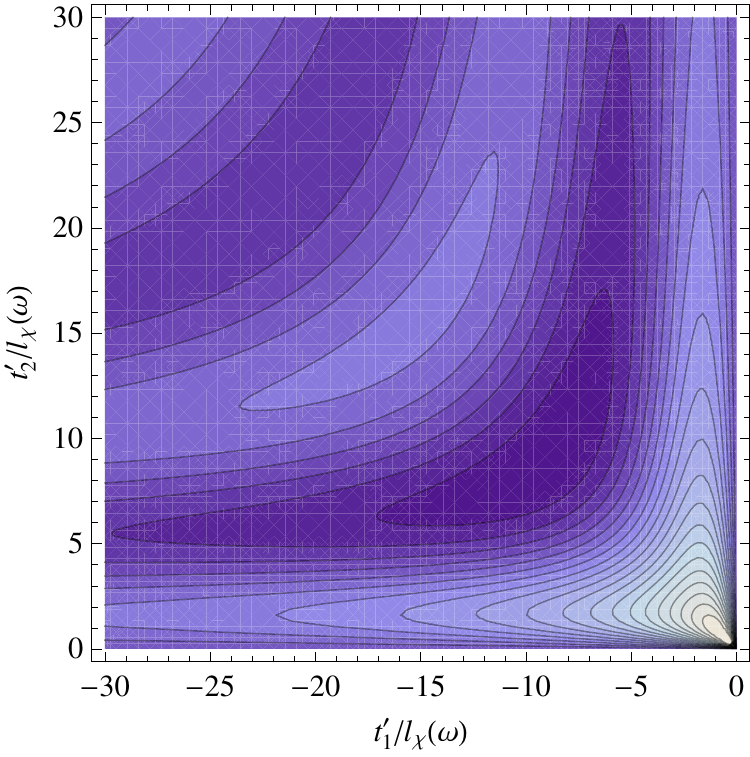}
 \caption{\label{fig:IBH-time-corr} Integrand of Eq.~(\ref{Iei}) at $\chi=30\gamma^{-1}$ [case of single scattering, corresponding to $\frac{dI_{\text{BH}}}{d\omega}(\gamma\chi)$]. The extended populated regions correspond to photon emissions along one of the external electron lines.}
\end{figure}

It will be instructive to begin with figuring out typical
contributing times for the simplest term
$\frac{dI_{\text{BH}}}{d\omega}(\gamma\chi)$. For visualization, let
us first of all plot the integrand of Eq.~(\ref{Iei}), which is depicted in
Fig.~\ref{fig:IBH-time-corr}. More quantitative conclusions require
scrutinizing the corresponding single integral (\ref{I0-w}). There,
typical $w(1-w)$, i.e., effectively, $\min\{w,1-w\}$, range from
$\sim1/\gamma^2\chi^2\ll 1$ to $\sim1$. According to
Eq.~(\ref{tau,deltatau}), that corresponds to typical
$\tau\sim\frac{2\gamma^2}{\omega\left[1+\gamma^2\chi^2w(1-w)\right]}$
ranging from $l_{\chi}(\omega)$ to $l_0(\omega)$. Invoking relations
$w=t'_2/\tau$, $1-w=-t'_1/\tau$, that can be expressed in terms of $t'_1$ and $t'_2$ as
\[
\frac{\max\left\{|t'_1|,t'_2\right\}}{l_0(\omega)}+\frac{\min\left\{|t'_1|,t'_2\right\}}{l_{\chi}(\omega)}\sim1.
\]
The strong inequality between the contributing times, arising when
$|t'_1|\sim l_0(\omega)\gg t'_2\sim l_{\chi}(\omega)$ or $|t'_1|\sim
l_{\chi}(\omega)\ll t'_2\sim l_0(\omega)$, reflects the fact that
photons are intensely emitted along the initial or final electron
direction. Among those, photons with coherence time $|t'_1|\sim
l_0(\omega)$ must be collinear to the initial electron (being intrajet),
whereas $ t'_2\sim l_{\chi}(\omega)$ then represents the formation
time for interjet photons (cf. Fig.~\ref{fig:Ang}), being
significantly different from that for intrajet photon formation. The
criterion of attributing the corresponding contribution to a jet is
its independence of the electron scattering angle, whereas interjet
radiation embodies all the dependence on this angle.

The interference integral (\ref{dIdomega=intdw}) may be treated in a
similar manner [first determining $w$, and next the times from
Eqs.~(\ref{tau,delta-tau-through-w}) and (\ref{tau,deltatau})], but
it requires different approximations in different spectral regions.

\subsection{High-$\omega$ domain. Intermediate electron contribution in Feynman gauge}

For the interference integral in Eq. (\ref{dIdomega=intdw}), first
consider the domain of high $\omega$, which is where the relatively
simple intermediate line contribution must build up. In this limit,
the cosine in Eq.~(\ref{dIdomega=intdw}) is rapidly oscillating.
Integrals from oscillatory functions are dominated by points of
stationary phase and end points of the integration interval
\cite{Olver}. In our case, there are no stationary phase points on
the real axis, whereas the lower end point essentially does not
contribute, because there $\cos \frac{\omega T}{2\gamma^2}
\left[\frac1w +\gamma^2\chi^2 \left(1-w\right)\right]$ oscillates
increasingly fast. Thus, the dominant contribution is brought by the
upper end point alone, with the leading terms there being
\begin{eqnarray}\label{gll-w-terms}
\int_{0}^{1}dw \cos\frac{\omega T}{2\gamma^2 w}\qquad\qquad\qquad\qquad\qquad\qquad\qquad\nonumber\\
-\int_{0}^{1} \frac{dw}{1+\frac1{\gamma^2\chi^2}-w}\cos \frac{\omega
T}{2\gamma^2} \left[\frac1w +\gamma^2\chi^2 \left(1-w\right)\right].
\end{eqnarray}
In the second integral (stemming from $\mathcal{I}_{ei}$), at
$\gamma\chi\gg1$ it is always legitimate replace in the phase $\frac1w\to1$,
since away from point $w=1$, this term plays minor role for any
$\omega T$. Besides that, the lower integration end point may be
replaced by $-\infty$, as long as $\frac{\omega T}{2\gamma^2}$ is
kept sizable. The result
\begin{eqnarray}\label{cos-slow}
\int_{0}^{1}dw \cos\frac{\omega T}{2\gamma^2 w}\qquad\qquad\qquad\qquad\qquad\qquad\qquad\nonumber\\
-\int_{-\infty}^{1} \frac{dw}{1+\frac1{\gamma^2\chi^2}-w}\cos
\frac{\omega T}{2\gamma^2} \left[1 +\gamma^2\chi^2
\left(1-w\right)\right] \nonumber\\
=g\left(\frac{\omega
T}{2\gamma^2}\right)\qquad\qquad\qquad\qquad\qquad\qquad\qquad\qquad
\end{eqnarray}
coincides with the intermediate electron line form factor
(\ref{g-def}).

It is worth noting that at $\omega\to\infty$, the leading
$\mathcal{O}(\omega^{-1})$ contributions from individual integrals
in (\ref{gll-w-terms}) mutually cancel, and the physical behavior $\sim\omega^{-2}$ is brought by the next-to-leading order contribution. In Sec.~\ref{subsec:2.1}, that property
was attributed the vector character of electromagnetic radiation,
via formation of the hollow cone angular distribution. Here, in the
Feynman gauge, the cancellation engages the trajectory-independent
part, represented by the first integral in (\ref{gll-w-terms}) or
(\ref{cos-slow}).

From the lhs of Eq.~(\ref{cos-slow}), one infers that at
$\frac{\omega T}{2\gamma^2}\gg1$, in both integrals typical
$w=t_2'/\tau$ are close to unity. That implies that
$t'_2\to\tau=t'_2-t'_1$, i.e., for $\mathcal{I}_{ei}$, the first
correlation time $t'_1\to-0$, tending to the first scattering point.
From Eq.~(\ref{tau-sim-Tw}) we also see that $\tau\approx T/w\to T$,
whence $t'_2\to T$, i.e., it tends to the second scattering point,
as is expectable physically. Finally, from the cosine factor of the
second integral in Eq.~(\ref{cos-slow}), yielding
\[
1-w\sim\frac2{\omega T\chi^2},
\]
and from
Eq.~(\ref{delta-tau-through-w}), we get
\begin{equation}
\delta\tau\sim l_0(\omega),
\end{equation}
which is natural from the collinear-collinear interference point of
view.\footnote{If one desires to estimate not only $\delta\tau$, but
variations of each of the contributing times, as well, it is
necessary to return to the original double-time representation
(\ref{Iei}), and linearize the argument of the sine in
Eq.~(\ref{Iei}) about point $t'_1=0$, $t'_2=T$:
\[
\sin\omega\left[\frac{\tau}{2\gamma^2}-\frac{\chi^2}2\frac{t'_1
t'_2}T\right]\simeq\sin\left[\frac{\omega}{2\gamma^2}(T+\delta
t'_2)-\frac{\omega\chi^2}2\delta t'_1\right].
\]
This shows that the extents of the contributing time regions are
unequal:
\begin{equation}
\delta t_1\sim l_{\chi}(\omega)\ll \delta t_2\sim l_{0}(\omega)\ll
T.
\end{equation}
For the $\mathcal{I}_{ie}$ part, vice versa, one would obtain
\begin{equation}
\delta t_1\sim l_{0}(\omega)\gg \delta t_2\sim l_{\chi}(\omega).
\end{equation}
Thus, at each end of the intermediate electron line, $l_{0}(\omega)$
and $l_{\chi}(\omega)$ enter on equal rights, although here, in
contrast to  $I_{\text{BH}}$, they are adjacent not to one, but to
different vertices.
} It is noteworthy that in spite of the dependence of one of the
$\delta t$'s on $\chi$, the resulting intermediate line spectral
contribution (\ref{Ci+si}) is $\chi$-independent, insofar as the
smallness of one of the time intervals, $\propto
\frac1{\gamma^2\chi^2}$, in the double time integral is exactly
compensated by the prefactor containing one power of
$\gamma^2\chi^2$ in the numerator. As for the phase, it is
independent of $\chi$, granted that at $t'_1\to0$,
$|\bm{r}(t'_2)-\bm{r}(t'_1)|\simeq vt'_2\simeq v\tau$. Physically,
the negligibility of the trajectory curvature, in spite of the
trajectory bending to a substantial angle, is chained to the fact
that at high $\omega$, this bending is felt only along a short
distance.

\subsection{Low-$\omega$ domain. `Radio' contribution and long time scales}

\begin{figure}
\includegraphics{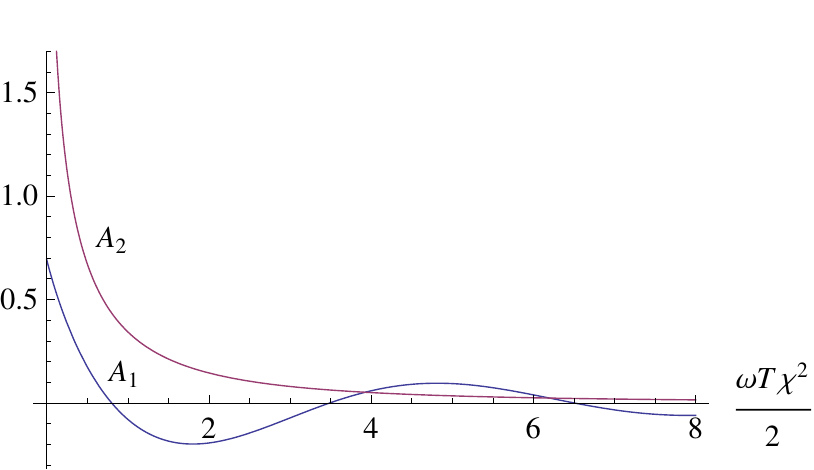}
 \caption{\label{fig:A12} Antenna form factors in the case of coplanar electron scattering through two equal angles.
 $A_1$ (blue curve) is given by Eq.~(\ref{A1-int}), and $A_2$ (red curve) by Eq.~(\ref{A2-int}).}
\end{figure}

Next, let us turn to the domain of low $\omega$. It must be remembered that the intermediate line contribution extends there, as well.
Approximation $w\approx1$ for the second term of
Eq.~(\ref{gll-w-terms}) remains valid even when $\omega\to0$ -- not
because of the influence of the cosine factor (which varies slowly
in the infrared limit), but due to the prefactor
$\frac1{1+\gamma^{-2}\chi^{-2}-w}$ peaking near the end point. The
only difference is that at $\omega T\chi^2/2\lesssim1$, the lower
end point in the second term cannot be replaced by $-\infty$, as in
Eq.~(\ref{cos-slow}). But to cope with the latter impediment, and extend
approximation (\ref{cos-slow}) to the low-$\omega$ region (where its
behavior will become logarithmic), it suffices just to subtract
therefrom the corresponding lower end point contribution
\begin{eqnarray}\label{add-subtr-integral}
-\int_{-\infty}^{0} \frac{dw}{1+\frac1{\gamma^2\chi^2}-w}\cos
\frac{\omega T}{2\gamma^2} \left[1 +\gamma^2\chi^2
\left(1-w\right)\right]\nonumber\\
\approx -\int_{-\infty}^{0} \frac{dw}{1-w}\cos \frac{\omega
T\chi^2}{2} \left(1-w\right).\qquad\qquad\quad
\end{eqnarray}
Combining (\ref{add-subtr-integral}) with the rest of the terms of
(\ref{dIdomega=intdw}), in the low-$\omega$ limit one gets radio contribution in the form
\begin{eqnarray}\label{FA-w}
\int_{-\infty}^{0} \frac{dw}{1-w}\cos \frac{\omega T\chi^2}{2}
\left(1-w\right)\qquad\qquad\qquad\qquad\qquad\qquad\nonumber\\
+\int_0^{\infty}\!
dw\Bigg(\frac1{w+\frac1{2\gamma^2\chi^2}}-\frac1{w+\frac1{\gamma^2\chi^2}}
\Bigg)\cos\frac{\omega T}{2\gamma^2}\!  \left(\frac1w
+\gamma^2\chi^2
\right)\nonumber\\
+\int_0^1 \frac{dw}{2-w}\cos \frac{\omega T\chi^2}{2}
\left(1-w\right).\qquad\qquad\qquad\qquad\qquad\quad
\end{eqnarray}
Here, in the second line we have neglected in the phase the small
term linear in $w$, given typical $w\lesssim\gamma^{-2}\chi^{-2}$,
and accordingly replaced the upper integration limit by infinity,
whereas in the third line, on the contrary, term $\sim w^{-1}$ in
the phase was neglected, since it affects the regular integrand only
in a small vicinity of the origin. Thereby one separates in
(\ref{FA-w}) the pure contribution from the end point $w=1$:
\begin{equation}\label{A2-int}
A_2\left(\frac{\omega T\chi^2}2\right)=\int_{-\infty}^1
\frac{dw}{2-w}\cos \frac{\omega T\chi^2}{2} \left(1-w\right),
\end{equation}
while the rest includes
\begin{subequations}\label{A1-int}
\begin{eqnarray}\label{A1-diffr}
A_1\left(\frac{\omega T\chi^2}2\right)=\int_0^{\infty}
dw\Bigg(\frac1{w+\frac1{2\gamma^2\chi^2}}-\frac1{w+\frac1{\gamma^2\chi^2}}
\Bigg)\nonumber\\
\times\cos \frac{\omega T}{2\gamma^2} \left(\frac1w +\gamma^2\chi^2
\right)\quad
\end{eqnarray}
and
\begin{equation}\label{A1-non-diffr}
\int_{-\infty}^{0} dw\left(\frac{1}{1-w}-\frac{1}{2-w}\right)\cos
\frac{\omega T\chi^2}{2} \left(1-w\right).
\end{equation}
\end{subequations}
A change of integration variable $\tilde w=-\frac1{\gamma^2\chi^2
w}$ here reveals equality of contributions (\ref{A1-diffr}) and
(\ref{A1-non-diffr}),
hence,
the result of integration in (\ref{FA-w}) amounts
$2A_1\left(\frac{\omega T\chi^2}2\right)+A_2\left(\frac{\omega
T\chi^2}2\right)$, in agreement with Eq.~(\ref{FA-def}). The
behavior of form factors (\ref{A2-int}) and (\ref{A1-int}), which
are now functions of a single variable, is illustrated in
Fig.~\ref{fig:A12}.

\begin{figure}
\includegraphics{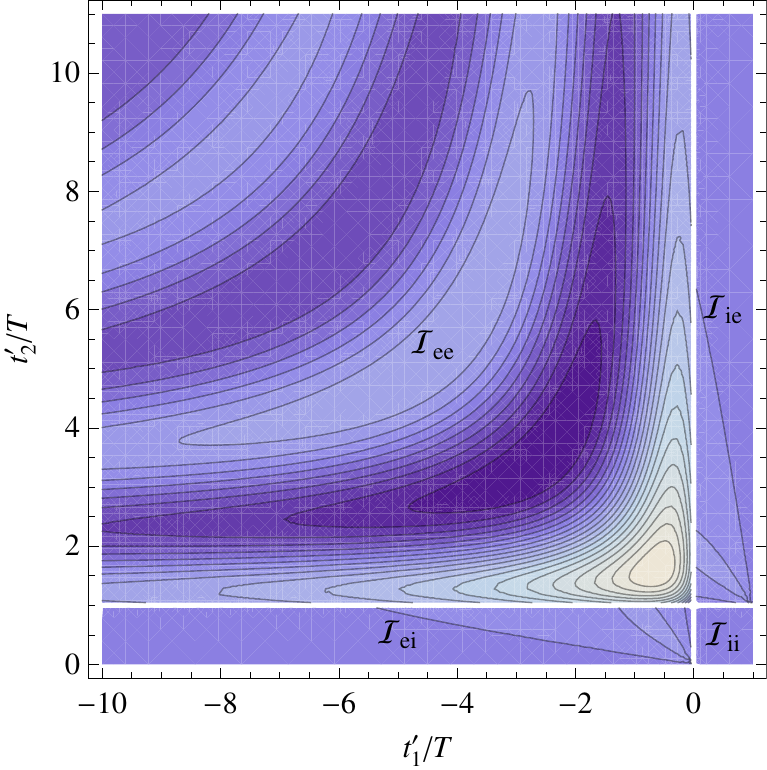}
 \caption{\label{fig:t1t2-lowomega} Integrand of Eq.~(\ref{t2tau-subtr}) at $\chi=30\gamma^{-1}$ and $\frac{\omega T}{2\gamma^2}=10^{-3}$.
 Most prominent is contribution $\mathcal{I}_{ee}$, similar to that of
 Fig.~\ref{fig:IBH-time-corr},  but now corresponding to $\frac{dI_{\text{BH}}}{d\omega}(2\gamma\chi)$.
 Also noticeable are contributions from $\mathcal{I}_{ei}$ and $\mathcal{I}_{ie}$,
 corresponding to parts (\ref{A2-int}) and (\ref{A1-non-diffr}) of the radio contribution,
 while part (\ref{A1-diffr}) is too broad to be captured by this figure.}
\end{figure}

For the evaluated radio part, again, it will be instructive first to
plot the integrand of Eq.~(\ref{t2tau-subtr}) in the $t'_1$, $t'_2$
plane (see Fig.~\ref{fig:t1t2-lowomega}), and with it in mind, analyze
Eqs.~(\ref{A2-int}), (\ref{A1-int}). From Eq.~(\ref{A1-diffr}), it
follows that $w\sim1/\gamma^2\chi^2\ll 1$, wherewith
Eq.~(\ref{tau,delta-tau-through-w}) gives
\begin{equation}\label{A1:tau-through-w}
\tau\sim\frac{T}{w}=\gamma^2\chi^2 T,
\end{equation}
and
\begin{equation}\label{A1:delta-tau-through-w}
\delta\tau\sim\frac{2\gamma^2}{\omega(1+\gamma^2\chi^2 w)}\sim
l_0(\omega)
\end{equation}
[which at $\omega T\chi^2\sim 1$ is commensurable with
(\ref{A1:tau-through-w})]. In terms of absolute times, that implies
\begin{equation}\label{diffr-typeII-timescales}
t_2,\, t'_2\sim w\tau\sim T, \qquad |t_1|\approx\tau\sim \gamma^2\chi^2T\gg t_2.
\end{equation}
The long extent of one of those times, just like in the case of
$I_{\text{BH}}$, indicates that the photon is formed within the
initial electron's proper field, and subsequently is stripped in the
electron scattering region.
Of course, there is also a cross-symmetric contribution, which has
been taken into account implicitly, by symmetry.

On the other hand, in integrals (\ref{A2-int}) and
(\ref{A1-non-diffr}) typical $w$ are of the order unity, entailing
\begin{equation}
|t_1|\sim t_2\sim T.
\end{equation}
That corresponds to the brightest spot in Fig.~\ref{fig:t1t2-lowomega}, but it is directly related only with $A_2$, since in artificial integral (\ref{A1-non-diffr}) all values of $w$ are unphysical (negative).

As we know from the preceding two sections, at $\omega
T\sim\chi^{-2}$, there arise spectral oscillations
$\sim\frac1{\omega T}\sin\frac{\omega T\chi^2}2$, related with
soft-collinear interference. Now we see that in integrals
(\ref{A1-diffr}) and (\ref{A1-non-diffr}) they stem from small time
ratios $w$ [in (\ref{A1-diffr}), formally -- from end point
$w=\infty$, but presently, that implies just
$\gamma^{-2}\chi^{-2}\ll
w\ll1$]. 
At the same time, contributing times for $A_2$ remain comparable
with $T$. That is the physical reason why $A_2$ does not need to be
supplemented by a form factor.

\subsection{Intermediate $\omega$ region: Decoherence and limits on ray optics}

With the increase of the photon frequency, terms $\propto w$ and
$w^{-1}$ in the phase in Eq.~(\ref{dIdomega=intdw}) eventually
become competing.
That first happens in the spectral region $\omega T\sim\gamma/\chi$, and implies that bending of the electron trajectory during the photon
formation process becomes essential (in accord with the notion of
ray optics established in the previous section).

As was already mentioned, there is no stationary phase point on the
real axis of $w$ (in contrast to the situation in
Sec.~\ref{subsec:imp-par}), so, in order to find the saddle point,
$w$ should be extended to the complex plane. But instead, it may suffice merely to note that typical contributing $w$ there are $\sim1/\gamma\chi\ll1$.
Simplifications are still possible then in the prefactors, which
reduce to pure power laws. Yet, there is a nontrivial term
$\frac1{1+\frac1{\gamma^2\chi^2}-w}\cos \frac{\omega T}{2\gamma^2}
\left[\frac1w +\gamma^2\chi^2 \left(1-w\right)\right]$, which
contributes not only at $w\sim1/\gamma\chi$, but also in vicinity of
the end point $w=1$, where it blows up due to the smallness of the
denominator. Those two different contributions in the same integral
may be just added:
\begin{eqnarray*}
\int_0^1dw\frac1{1+\frac1{\gamma^2\chi^2}-w}\cos \frac{\omega
T}{2\gamma^2} \left[\frac1w +\gamma^2\chi^2
\left(1-w\right)\right]\quad\\
\simeq \int_0^1dw\frac1{1+\frac1{\gamma^2\chi^2}-w}\cos \frac{\omega
T}{2\gamma^2} \left[1 +\gamma^2\chi^2
\left(1-w\right)\right]\\
+\int_0^1dw\cos \frac{\omega T}{2\gamma^2} \left[\frac1w
+\gamma^2\chi^2 \left(1-w\right)\right].\qquad\qquad\quad
\end{eqnarray*}
In the first (upper end point) term, the lower limit may be replaced
by $-\infty$, and along with term $\int_{0}^{1}dw \cos\frac{\omega
T}{2\gamma^2 w}$, it constitutes the omni-present intermediate
electron line contribution (\ref{cos-slow}). The second (inner-point) term combines
with the rest in Eq.~(\ref{dIdomega=intdw}) to give, under
conditions $\gamma^{-2}\chi^{-2}\ll w\ll1$,
\begin{eqnarray}\label{int-for-fast}
\int_{0}^{1}dw \Bigg(\frac1{w+\frac1{2\gamma^2\chi^2}}-\frac1{w+\frac1{\gamma^2\chi^2}}+\frac1{2-w}-1\Bigg)\qquad\qquad\,\nonumber\\
\times\cos \frac{\omega T}{2\gamma^2}\!\left[\frac1w +\gamma^2\chi^2
\left(1-w\right)\right]\nonumber\\
\simeq\frac1{2}\int_0^{\infty}
dw\left(\frac{1}{\gamma^2\chi^2w^2}-1\right)\cos \frac{\omega
T}{2\gamma^2}\!\left[\frac1w +\gamma^2\chi^2
\left(1-w\right)\right]\nonumber\\
=-\frac2{\gamma\chi}\sin\frac{\omega
T\chi^2}{2}K_1\left(\frac{\omega
T\chi}{\gamma}\right).\qquad\qquad\qquad\qquad\qquad\quad
\end{eqnarray}
That is exactly asymptotics (\ref{interm-omega-K1}). Adding up intermediate line and $F_{\perp}$-modulated radio
contributions, as in previous sections, ultimately recovers nondipole
decomposition (\ref{dIdomega-sum-formfactors}).

To estimate the relevant contributing times, let us note, again,
that $\tau\to T/w$ [Eq. (\ref{tau-sim-Tw})], wherefore $t_2\equiv w\tau/2\to T/2$, $t'_2\equiv w\tau\to T$, confirming that the second correlating time tends to the second
scattering vertex, in the spirit of Fig.~\ref{fig:Diagr}(b). As
for $t_1$, estimate $w\sim1/\gamma\chi$ implies
\begin{equation}\label{tau-sim-gammachi}
\tau\sim\gamma\chi T.
\end{equation}
That is again consistent with the ray optic notions: longitudinal
scale (\ref{tau-sim-gammachi}) equals the (fixed) transverse scale
$T\chi$ divided by the natural jet collimation angle $\gamma^{-1}$.
Hence,
\begin{equation}\label{times-for-SCI}
\tau,|t_1|\sim \gamma\chi T\gg t_2\sim T.
\end{equation}
Of course, there also exists a symmetric contribution $|t_1|\sim
T\ll t_2\sim  \gamma\chi T$, which had been taken into account
implicitly by doubling $\mathcal{I}_{ei}$ and $\int_0^1 dw\ldots$ in
$\mathcal{I}_{ee}$.


Strong inequality (\ref{times-for-SCI}) between the formation time
scale for interfering photons is in accord with the causal origin of
the factorization property: One of the two interfering components of
the electromagnetic wave forms up long before or long after another (which forms fast), wherefore they are causally
disconnected. At the same time, compared to the impact parameter
approach, the notion of the ray of light along which the
interference builds up is more uncertain here, because one of the
correlating times is broadly distributed ($\delta \tau\sim\tau$).
Thus, in the present problem, even in the domain of its best
applicability, the notion of ray optics is limited: The ray is well
defined within the double scattering region, but cannot be extended
down to the emission point. That makes the photon formation process
in the present case akin to diffraction.

The exponential falloff here appears to be due to a superficially different reason --
decoherence: Formally, the integrand in Eq.~(\ref{int-for-fast})
assumes a saddle point on the imaginary axis of $w$, and it is the
value of $e^{i\frac{\omega T}{2\gamma^2} \left[\frac1w
+\gamma^2\chi^2 \left(1-w\right)\right]}$ in this point which
converts to the exponentially decreasing factor. But physically, it
is due to phase fluctuations (typical values of the $w$-dependent
terms) $\frac{\omega T}{2\gamma^2w},\frac{\omega
T\chi^2}{2}w\sim\frac{\omega T\chi}{2\gamma}\lesssim 1$, which grow
with $\omega$, and progressively destroy the stability of the phase.


Although the mechanism of attenuation of low-$\omega$ spectral
oscillations looks different in different frameworks, there is a
noteworthy universal relation between the indeterminacies of the
photon formation time and the transverse screening scale:
\begin{equation}\label{l_perp-deltal_coh}
\chi l_{\perp}^{-1}=\delta l_{\text{f}}^{-1}.
\end{equation}
Here $\delta l_{\text{f}}^{-1}$ is the indeterminacy of the
reciprocal coherence length considered as a function of $\theta$ [Eq.~(\ref{coh-length-theta})],
or function of $w$,
\begin{equation}
l_{\text{f}}^{-1}=\frac{\omega }{2\gamma^2} \left[\frac1w
+\gamma^2\chi^2 \left(1-w\right)\right],
\end{equation}
with respect to typical indeterminacies $\delta\theta=2\chi/\gamma$
or $\delta w\sim1/\gamma\chi$.

In those basic considerations, we could not bring out all the
aspects of photon formation in the present process, so they may deserve additional investigation in the future.




\section{Experimental feasibility}\label{sec:feasibility}

To accomplish the study of radiation at double electron scattering,
it may also be expedient to discuss prospects for its experimental
realization. Promising candidates for prompt deflection of
relativistic particles to angles in excess of $\gamma^{-1}$ are thin
crystals. There are several known crystal-assisted deflection
mechanisms: channeling in a bent crystal \cite{channeling-steering},
volume reflection in a bent crystal \cite{VR-Taratin}, and mirroring
in a straight ultrathin ``half-wavelength" crystal
\cite{half-wavelength-crystal}. The acceptance to a stable
channeling mode in practice may be insufficiently high (see, e.g.,
\cite{Bagli}), whereas for mirroring in a ``half-wavelength"
crystal, the relative spread in deflection angles must be sizable
due to the impact parameter dependence. Volume reflection is not
beset by such deficiencies, so we examine it in the first place.

Volume reflection develops over a length $\Delta z_{\text{VR}}\sim
R\theta_c$, and for the case of positively charged particles (for
which it works somewhat better) leads to deflection to an angle
$\chi\approx \frac{\pi}2\theta_c$ \cite{Bond-VR}, where
$\theta_c=\sqrt{2V_0/E}\sim\sqrt{50\text{ eV}/E}$ is the critical
channeling angle. The intrinsic relative spread of the deflected
beam is $\Delta\chi/\chi\sim 2R_c/R$, where
$R_c=E/|F_{\max}|=0.2\text{ m}\frac{E}{\text{GeV}}$ is the critical
radius, and $R$ the crystal bending radius, which must be in excess
of $4R_c$.

Assuming positron energy $E=500$ GeV, which can become available in
the foreseeable future, such a positron can be deflected to an angle
$\chi\approx 1.5\theta_c=15\,\mu\text{rad}=15\gamma^{-1}$ within a
length $\Delta z_{\text{VR}}\approx\frac{R}{R_c}10\mu\text{m}\sim
0.1$ mm. The photon energy
$\omega\sim\frac1{T\chi^2}=1\frac{\text{MeV }\text{mm}}{T}$ will
belong to soft gamma range $\omega\sim1$ MeV provided the gap width
amounts $T\sim 1$ mm. As long as this is well in excess of $\Delta
z_{\text{VR}}$, the suggested setup should be feasible. At that, the
additional angular spread due to incoherent multiple scattering on
atomic nuclei in the crystal will be minor. An issue at such high an
energy can be synchrotron radiation background from steering and
focusing magnets, but it will be common for all the forward physics
problems. Other mechanisms of crystal deflection demand more
dedicated calculations.

Another option may be to utilize for deflection amorphous foils
equipped by a position sensitivity system (charged particle
tracking) enabling reconstruction of the electron trajectory and
thereby selection of events of double hard scattering through
prescribed angles. An issue therewith is that at momentum transfers
$E\chi=m_e\gamma\chi\gtrsim20m_e\sim10$ MeV, it may be important to
take into account inner structure of atomic nuclei. If such a setup
nonetheless proves feasible, the lower bound on the electron beam
energy could be relaxed. Condition $\gamma\gtrsim\sqrt{\omega T/2}$
(necessary for probing intermediate electron line resonances) with
$\omega\sim1$ MeV (to ensure transparency of both targets) and
$T>0.2$ mm implies $E=m_e\gamma>15$ GeV. It has been actually tested
in CERN, without electron tracking, at $E\sim200$ GeV and
$\omega\sim1$ GeV \cite{NA63-plans}. As for condition
$\chi^{-1}\sim\gamma/30\gtrsim\sqrt{\omega T/2}$ necessary for
testing radio resonances, it can be made compatible with CERN SPS
energies $E\sim200$ GeV for similar parameters $\omega\sim1$ MeV and
$T\sim0.2$ mm.

If any kind of electron hard rescattering and observation of
interference in the accompanying radiation will be realized, it
would open prospects for experimental tests of coherence phenomena
similar to those for quantum field theory jets. Let us remind that
it is actually the coherence that distinguishes gauge field theory
jets from purely random parton cascading \cite{Dokshitzer}. At that,
the notion of jets is usually associated with angular distributions,
so it would be desirable as well to measure angular distributions of
radiation like those in Fig.~\ref{fig:Ang}. Simultaneous measurement
of photon energy and (small) emission angle is a challenge similar
to that in gamma astronomy, which stimulates development of
pixellated detector arrays \cite{pixel-gamma-detectors}.

Finally, the electron deflection can be carried out by means of
magnet deflectors in vacuum, but since magnet dimensions are always
formidable, gap $T$ must be large, too, and correspondingly, the
interesting radiation will not fall into gamma range. Experiments in
optical region, including measurements of radiation angular
distributions, had been undertaken some time ago \cite{Nikitin}.
Under those conditions, though, one generally has to regard
near-field effects (see, e.g., \cite{Geloni}).

\section{Summary}\label{sec:summary}

The principal prediction of the present paper is that when an
electron is subjected to a double hard scattering through definite
angles, the spectrum of the emitted radiation exhibits oscillations
in two regions, reflecting manifestations of two coherence lengths:
free [$l_0(\omega)$] and electron scattering angle dependent
[$l_{\chi}(\omega)$].


The underlying reason for such an oscillatory behavior is the
interplay of two categories of photons: those formed along straight
parts of the electron's trajectory, with formation scale
$l_0(\omega)$, and those emerging from relatively small vicinities
of the trajectory break points, and forming at scale
$l_{\chi}(\omega)$. Radiation of the first type is narrowly
collimated along parent electron lines (intrajet, or collinear
radiation), whereas that of the second type is broadly distributed
in between the radiation jets (interjet radiation). Fainter angular
distribution of the latter ($\sim\gamma^{-2}\chi^{-2}$) is
compensated by its wider occupied phase space
($\sim\gamma^2\chi^{2}$), so in the angle-integral spectrum those
contributions are comparable.



Spectral oscillations, persisting in spite of integration over all photon emission angles, arise when there are two interfering radiation components. At least one among them
must be of collinear type, because by virtue of its natural narrow
collimation properties, it can carry a well-defined phase. The
second interfering component then must be emitted along the same
direction. For the certainty of the phase, besides that, both
components must have approximately equal impact parameters, i.e.,
effectively belong to the same ray in position space within the
scattering region. One should then distinguish two kinds of
interference geometries:
\begin{enumerate}
  \item Interference between electromagnetic waves emitted from opposite ends of the intermediate segment of the
  electron's trajectory close to the direction of its velocity,
  and having small impact parameters [see Fig.~\ref{fig:Diagr}(a)].
  Both interfering waves here are collinear to the same electron line.
  This type of interference was discussed in
  \cite{Blankenbecler,Zakharov,BK-structured,Bondarenco-Shulga,synchr-rad-straight-section,Geloni}.

  \item Interference between electromagnetic waves, one of which is emitted from one of the external electron lines and keeps collinear to it,
  and another one (interjet), from the opposite vertex. Those waves propagate nearly parallel to the corresponding external electron line,
  at an impact parameter such that they pass through the opposite vertex [see Fig.~\ref{fig:Diagr}(b)].
  The photon formation length here amounts $l_{\chi}(\omega)$ -- in spite of formation length for one of the waves being $l_0(\omega)$,
  the coherence length equals the smallest between the two.
\end{enumerate}

The formal realization of the scale separation property is nondipole
spectral decomposition (\ref{dIdomega-sum-formfactors}). Therein,
each term or factor depends on $\omega$ at its intrinsic scale, and
contains appropriate approximations, but formally extends through
the whole $\omega$ range. Interfering radiation from the
intermediate electron line is associated with term $g\left({\omega
T}/{2\gamma^2}\right)$ given by Eq.~(\ref{g-def}). The rest of the
terms are ``radio" contributions factorizing into the quasiantenna
[Eqs.~(\ref{A1-def}), (\ref{A2-def})] and the suppressing proper field form factors [Eq.~(\ref{Fj-def})]. 
The latter form factors furnish the exponential damping of the soft
spectral oscillations with the increase of $\omega$ due to
localization of the interfering waves at a nonzero impact parameter
$T\chi$, and due to decrease of the intrajet photon impact parameter
distribution (on a scale given by the transverse coherence length),
or, equivalently, due to fluctuations of the longitudinal coherence
length [Eq.~(\ref{l_perp-deltal_coh})]. Taken apart, soft and hard
terms in the spectral density diverge at $\omega\to0$
logarithmically, $g,r\sim\pm\ln\frac1{\omega}$ (cf. \cite{Goldman}),
but their sum is finite.

It is likely that similar decomposition and factorization properties
will prove relevant also in other problems involving continuous
targets with sharp boundaries.

\subsection*{Acknowledgements}

M.V.B. is grateful to A.V. Shchagin for bringing to his attention
Ref.~\cite{synchr-rad-straight-section}. This work was supported in
part by the National Academy of Sciences of Ukraine (Project No.
CO-1-8/2016) and the Ministry of Education and Science of Ukraine
(Project No. 0115U000473).

\appendix

\section{Derivation of representation (\ref{t2tau-subtr})}
Representation (\ref{dIdomega-through-amp}) in form of a double time
integral
\begin{eqnarray}\label{dIdomega-through-double-time-int}
\frac{dI}{d\omega}=\left(\frac{e\omega}{2\pi}\right)^2\!\int d^2n
\iint_{-\infty}^{\infty}dt_1dt_2\left[\bm{n}\times\bm{v}(t_1)\right]\cdot\left[\bm{n}\times\bm{v}(t_2)\right]\,\nonumber\\
\times e^{i\omega (t_1-t_2)-i\bm{k}\cdot
\left[\bm{r}(t_1)-\bm{r}(t_2)\right]}\qquad
\end{eqnarray}
allows exact integration over radiation angles. To this end, $\int
d^2n$ must be performed prior to integration over $t_1$ and $t_2$.
It should be minded that time integrals in
(\ref{dIdomega-through-double-time-int}) are not absolutely
convergent, so change of the integration order compared to
Eq.~(\ref{dIdomega-through-double-time-int}) must be done carefully. Problems
arise in the limit $t_2\to t_1$, where the angular integral from the
oscillatory exponential becomes singular. In particular, it may be
necessary to treat the emerging singular function there as an
improper one (a distribution).

Integration in (\ref{dIdomega-through-double-time-int}) can be
simplified by employing gauge invariance to reduce the power of
$\bm{n}$ in the preexponential factor. Rewriting
\begin{equation}\label{nvnv=vv-nvnv}
\left[\bm{n}\times\bm{v}(t_1)\right]\cdot\left[\bm{n}\times\bm{v}(t_2)\right]=v_i(t_1)v_k(t_2)(\delta_{ik}-n_i
n_k),
\end{equation}
one can replace the photon polarization density matrix
$\delta_{ik}-n_i n_k$ by that in the covariant (Feynman) gauge,
proportional to the metric tensor
$g_{\mu\nu}=\text{diag}(1,-1,-1,-1)$ in Minkowski space-time, and
not involving $\bm{n}$:
\begin{eqnarray}\label{Feynman-gauge}
dt_1 dt_2 v_i(t_1)v_k(t_2)(\delta_{ik}-n_i n_k) \to -ds_1 ds_2 u_{\mu}(t_1)u_{\nu}(t_2)g_{\mu\nu}\nonumber\\
=dt_1 dt_2\left[\bm{v}(t_1)\cdot \bm{v}(t_2)-1 \right],\nonumber\\
\end{eqnarray}
where $u_{\mu}=dr_{\mu}/ds=\gamma (1,\bm{v})$,
$ds=\sqrt{dt^2-d\bm{r}^2}=dt/\gamma$. The validity of form
(\ref{Feynman-gauge}) can equally well be justified via integration
by parts in the second term of (\ref{nvnv=vv-nvnv}):
\begin{eqnarray}\label{ident-byparts}
\int_{-\infty}^{\infty}dt \bm{n}\cdot\bm{v}(t) e^{i\left\{\omega t-\bm{k}\cdot \bm{r}(t)\right\}}=\frac{i}{\omega} \int_{-\infty}^{\infty}dt e^{i \omega t}\frac{\partial}{\partial t} e^{-i\bm{k}\cdot \bm{r}(t)}\nonumber\\
=-\frac{i}{\omega} \int_{-\infty}^{\infty}dt e^{-i\bm{k}\cdot
\bm{r}(t)} \frac{\partial}{\partial t} e^{i \omega
t}=\int_{-\infty}^{\infty}dt e^{i\omega t-i\bm{k}\cdot
\bm{r}(t)}\quad
\end{eqnarray}
for each of the times $t_1$, $t_2$. The change of the gauge as a
result of integration by parts is the fundamental property of
electrodynamics \cite{Jackson}.

Inserting (\ref{Feynman-gauge}) to
(\ref{dIdomega-through-double-time-int}) and making simplifications
pertinent to the ultrarelativistic limit, leads to
\cite{Akh-Shul-UFN1982}
\begin{eqnarray}\label{AS-t1t2-v1-v2}
\frac{dI}{d\omega}= -\left(\frac{e\omega}{2\pi}\right)^2\int d^2n \iint_{-\infty}^{\infty}dt_1dt_2\qquad\qquad\qquad\qquad \nonumber\\
\times \left\{\gamma^{-2}+\frac12\left[\bm{v}(t_2)-
\bm{v}(t_1)\right]^2\right\} e^{i\omega (t_1-t_2)-i\bm{k}\cdot
\left[\bm{r}(t_1)-\bm{r}(t_2)\right]}.\nonumber\\
\end{eqnarray}
Here it was presumed that
$\bm{v}^2(t_1)=\bm{v}^2(t_2)=1-\gamma^{-2}$ is time independent
(otherwise $\gamma^{-2}$ must be replaced by
$\frac12\left[\gamma^{-2}(t_1)+\gamma^{-2}(t_2)\right]$).

Next, we employ the symmetry between $t_1$ and $t_2$ to write
$\iint_{-\infty}^{\infty}dt_1dt_2\ldots=2\mathfrak{Re}\int_{-\infty}^{\infty}dt_2\int^{t_2}_{-\infty}dt_1\ldots$,
and note that integral
\[
\int d^2n e^{i\bm{k}\cdot
\left[\bm{r}(t_2)-\bm{r}(t_1)\right]}=\pi\int_0^{\infty}
dn_{\perp}^2 e^{i\omega \left(1-n_{\perp}^2/2\right)
\left|\bm{r}(t_2)-\bm{r}(t_1)\right|}
\]
will converge absolutely provided we replace
$\left|\bm{r}(t_2)-\bm{r}(t_1)\right|\to
\left|\bm{r}(t_2)-\bm{r}(t_1)\right|-i\epsilon$, where
$\epsilon\to+0$. The integration then gives \cite{BKS}
\begin{eqnarray}\label{dIdomega-t2-t1+iepsilon}
\frac{dI}{d\omega}= -\omega\frac{e^2}{\pi}\int_{-\infty}^{\infty}dt_2\int^{t_2}_{-\infty}dt_1 \left\{\gamma^{-2}+\frac12\left[\bm{v}(t_2)- \bm{v}(t_1)\right]^2\right\}\nonumber\\
\times \mathfrak{Im}\frac{1}{t_2-t_1-i\epsilon} e^{-i\omega
\left[t_2-t_1-\left|\bm{r}(t_2)-\bm{r}(t_1)\right|\right]},\qquad
\end{eqnarray}
where we replaced in the preexponential factor
$\left|\bm{r}(t_2)-\bm{r}(t_1)\right|\approx t_2-t_1$, while in the
phase factor such a replacement is generally not justified.

\begin{figure}
\includegraphics{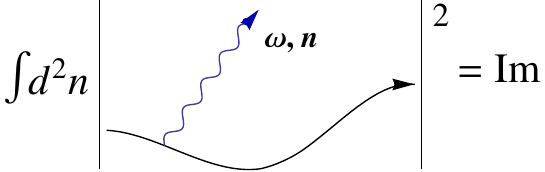}\includegraphics{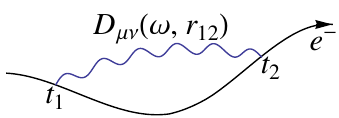}
 \caption{\label{fig:loop-diagram} Graphical illustration of Eq.~(\ref{dIdomega-photon-proparator}).}
\end{figure}

The meaning of formula (\ref{dIdomega-t2-t1+iepsilon}) becomes obvious when written covariantly as
\begin{eqnarray}\label{dIdomega-photon-proparator}
\frac{dI}{d\omega}= \omega\frac{e^2}{\pi}\int_{-\infty}^{\infty}ds_2\int^{s_2}_{-\infty}ds_1 u_{\mu}(t_1)u_{\nu}(t_2)\qquad\nonumber\\
\times\mathfrak{Im} e^{-i\omega
(t_2-t_1)}D_{\mu\nu}\left(\omega,\left|\bm{r}(t_2)-\bm{r}(t_1)\right|\right),
\end{eqnarray}
where
\[
D_{\mu\nu}(\omega,r)=-\frac{g_{\mu\nu}}{r-i\epsilon} e^{i\omega r}
\]
is the photon propagator in Feynman gauge and frequency-position
representation \cite{BLP} (appropriately regularized at $r=0$, which
would have no effect in quantum electrodynamics, but is essential in
classical). Equation~(\ref{dIdomega-photon-proparator}) expresses
nothing but the unitarity relation (cf., e.g., \cite{BLP}) between
the angle-integral real photon emission probability $\frac1{\hbar
\omega}\frac{dI}{d\omega}$ and the imaginary part of a virtual
photon propagator inserted between two points on the electron
trajectory -- as is graphically illustrated in
Fig.~\ref{fig:loop-diagram}. Notation
(\ref{dIdomega-photon-proparator}) is gauge invariant, holding in
any gauge for the photon propagator, but the use of Feynman gauge is
arguably the simplest.

The effect of infinitesimal term $-i\epsilon$ in the denominator of $D_{\mu\nu}$ is
that
\begin{eqnarray}\label{delta-fn}
-\mathfrak{Im}\frac{1}{t_2-t_1-i\epsilon}
e^{-i\omega \left[t_2-t_1-\left|\bm{r}(t_2)-\bm{r}(t_1)\right|\right]}\qquad\qquad\quad\nonumber\\
\underset{\epsilon\to+0}\to\frac{\sin\omega
\left[t_2-t_1-\left|\bm{r}(t_2)-\bm{r}(t_1)\right|\right]}{t_2-t_1}-\pi\delta(t_2-t_1).
\end{eqnarray}
Here, since the singularity point of the emerging delta function
falls onto the integration domain edge in
Eq.~(\ref{dIdomega-t2-t1+iepsilon}), due to the symmetry between $t_1$
and $t_2$, the contribution from the delta function must be regarded
as \emph{halved}. Owing to the last term, the radiation spectrum
vanishes for a uniform and rectilinear electron motion.

In practice, it may be convenient to replace the delta function (the
instantaneous term) in (\ref{delta-fn}) by a regular function
producing an identical effect. Customarily, it is written as
\begin{eqnarray}\label{mathcal-K}
\frac{dI}{d\omega}
=\omega\frac{e^2}{\pi}\int_0^{\infty}\!\frac{d\tau}{\tau}\int_{-\infty}^{\infty}\! dt_2\Bigg(\!\! \left\{\!\gamma^{-2}+\frac12\left[\bm{v}(t_2)- \bm{v}(t_2-\tau)\right]^2\right\}\nonumber\\
\qquad\qquad\qquad\times \sin \omega \left[\tau-\left|\bm{r}(t_2)-\bm{r}(t_2-\tau)\right|\right]\nonumber\\
\qquad\qquad\qquad\qquad\qquad-\gamma^{-2}
\sin\mathcal{K}\tau\Bigg)\qquad\qquad
\end{eqnarray}
with $\mathcal{K}\to\infty$, or, since
$\int_0^{\infty}\frac{d\tau}{\tau}\sin\mathcal{K}\tau=\frac{\pi}{2}$
is actually $\mathcal{K}$-independent, in form (\ref{t2tau-subtr}).
The advantage of the latter form is that for a uniformly and
rectilinearly moving charge, the integrand rather than only the
whole integral turns to zero. (Yet, since the integrand becomes
decreasing as $|t_2|\to\infty$, it affords one to interchange the
order of integrations.) Equation~(\ref{t2tau-subtr}) is the
subtracted Blankenbecler-Drell formula \cite{Blankenbecler-Drell},
which was derived here without introducing the ``vacuum" term by
hand.

It is also worth noting that representation (\ref{mathcal-K}) with
$\mathcal{K}=2\omega$ can be obtained directly if one integrates not
only over typical small photon emission angles, but over the full solid
angle \cite{Akh-Shul-UFN1982}. Then, $\sin2\omega\tau$ is associated
with ``backward" radiation, which may be physically negligible, but
is suitable for regularizing the integral.

\end{document}